\documentclass[journal,comsoc]{IEEEtran}
\usepackage{graphicx}
\usepackage{mathtools}
\usepackage{amsmath,amsfonts,amssymb}
\usepackage{url}
\usepackage{makecell}
\usepackage{float}
\usepackage{subfig}
 \usepackage[table]{xcolor}
\usepackage{cite}
\usepackage{flushend}
\usepackage{enumitem}
\usepackage{pdfpages}
\usepackage{booktabs, multicol, multirow}
\newcommand{\beq}{\begin{equation}}
\newcommand{\eeq}{\end{equation}}
\newcommand{\beqa}{\begin{eqnarray}}
\newcommand{\eeqa}{\end{eqnarray}}
\newcommand{\SubItem}[1]{
	{\setlength\itemindent{15pt} \item[-] #1}
}
\DeclareMathOperator*{\argmax}{arg\,max}
\DeclareMathOperator*{\argmin}{arg\,min}

\usepackage[cmintegrals]{newtxmath}

\begin{document}

\title{Experimental Review of Neural-based approaches for Network Intrusion Management}

\author{Mario~Di~Mauro,~\IEEEmembership{Member,~IEEE,}
	Giovanni~Galatro,
	Antonio~Liotta,~\IEEEmembership{Senior Member,~IEEE}
	\IEEEcompsocitemizethanks{\IEEEcompsocthanksitem M. Di Mauro and G. Galatro are with the Department of Information and Electrical Engineering and Applied Mathematics (DIEM), University of Salerno, 84084, Fisciano, Italy (E-mail: mdimauro@unisa.it,giovanni.galatro@gmail.com).
		\newline
		
		A. Liotta is with the Free University of Bozen-Bolzano, Italy (E-mail: Antonio.Liotta@unibz.it)
		
	}
}

\maketitle

\begin{abstract}
	The use of Machine Learning (ML) techniques in Intrusion Detection Systems (IDS) has taken a prominent role in the network security management field, due to the substantial number of sophisticated attacks that often pass undetected through classic IDSs. These are typically aimed at recognizing attacks based on a specific signature, or at detecting anomalous events. However, deterministic, rule-based methods often fail to differentiate particular (rarer) network conditions (as in peak traffic during specific network situations) from actual cyber attacks. 
	In this paper we provide an experimental-based review of neural-based methods applied to intrusion detection issues. Specifically, we \textit{i)} offer a complete view of the most prominent neural-based techniques relevant to intrusion detection, including deep-based approaches or weightless neural networks, which feature surprising outcomes; \textit{ii)} evaluate novel datasets (updated w.r.t. the obsolete KDD$99$ set) through a designed-from-scratch Python-based routine; \textit{iii)} perform experimental analyses including time complexity and performance (accuracy and F-measure), considering both single-class and multi-class problems, and identifying trade-offs between resource consumption and performance.
	Our evaluation quantifies the value of neural networks, particularly when state-of-the-art datasets are used to train the models. This leads to interesting guidelines for security managers and computer network practitioners who are looking at the incorporation of neural-based ML into IDS.  
\end{abstract}


\section{Introduction}
\label{intro}

Over the past few years, the adoption of Machine Learning (ML) techniques to the field of network security has become prominent \cite{granville1,granville2}. This is mainly due to the possibility of tackling a range of ever more sophisticated attacks able to circumvent security-based systems which rely on classic features inspection
(e.g. port-control, signature-based flow detection, IPs black-listing, etc.). 
It is useful to recall that, especially in encrypted traffic analysis, the difference between \textit{deterministic} and \textit{stochastic} features is crucial. Deterministic ones pertains to ``static'' information embodied in security protocols such as TLS (e.g. record length, handshake types, cipher suites, etc.) \cite{cisco-nids}, or IPSec (ISAKMP SPI Initiator/Responder, payload length, etc. ). On the other hand, stochastic features exploit the probabilistic nature (hard to hide in encrypted flows as well) of some traffic characteristics (e.g. the distribution of inter-arrival times). On behalf of ML-based techniques, it becomes quite straightforward to manage stochastic features which, coupled with the deterministic ones, allow to characterize as accurately as possible an encrypted data flow.  
	
Moreover, new network intrusion detection systems (NIDS) \cite{gaspary1,gaspary2} can actually interact with ML-based engines in order to learn the statistical features that characterize the various traffic flows and, in turn, classify them according to specific performance/time efficiency trade-offs.
Among various possibilities of interaction, the Simple Network Management Protocol (SNMP) offers a flexible and standardized way to collect traffic data from a number of agents, by relying on Management Information Base (MIB) objects which provide information about some features to be managed  \cite{cerroni1,cerroni2,cerroni3}.

This notwithstanding, one of the biggest problems is to deal with the jungle of ML techniques which, depending on the underlying strategy, can exhibit completely different performance when applied in the intrusion detection field.

%There is, however, a vast variety of ML algorithms, often based on completely different principles, which makes it very difficult to carry out comprehensive comparisons. 

In fact, most attempts at surveying ML-based traffic classification problems have resulted in non-homogeneous comparisons and often unfair outcomes. 
Another flaw found in existing literature concerns the choice of a valid dataset. Most of studies, in fact, rely on the obsolete KDD$99$ dataset \cite{kdd-dataset}, or its evolved version NSL-KDD \cite{nsl-kdd}. These datasets either do not contemplate essential features of modern data traffic (e.g. voice, video), or contain outdated signatures of network attacks \cite{icissp-dataset3}.

In this paper we address the aforementioned shortcomings, providing the following main contributions. \textit{i)} We survey neural-based techniques, which have recently taken prominence thanks to deep-based methods, in the specific context of traffic classification problems; we provide a critical comparison of algorithms that share a common paradigm (e.g. deep neural networks, linear vector quantization, etc.), including also techniques that are not typically  applied to intrusion detection, such as weightless neural networks, whose results are quite unexpected. \textit{ii)} We go beyond a traditional survey, providing an experimental-based assessment; we take into account novel traffic datasets (such as CIC-IDS-2017/2018), where both single-class cases (namely benign vs. malign) and multi-class cases (namely benign vs. malign$_1$ ... vs. malign$_k$) are considered; we perform both performance analysis (through accuracy/F-measure figures) and time-complexity analysis. 

The paper is organized as follows. Section \ref{sec:rw} presents an {\em excursus} of related literature and, by means of a comparative table, we highlight differences and commonalities with other surveys. In Sect. \ref{sec:nn} we review the most credited neural-based methods. Section \ref{sec:dataset} presents details about the novel considered datasets, where traffic features are grouped according to a similarity criterion. In Sect. \ref{sec:experim} we provide an experimental-based comparison, where different neural-based techniques are juxtaposed through performance and time-complexity analyses. Finally, Section \ref{sec:conclusion} concludes this work by also indicating some promising research directions.    

\section{Related Work on ML techniques applied to network intrusion detection}
\label{sec:rw}

ML-based intrusion detection is becoming attractive for a variety of network environments including service providers \cite{classif_dimauro}, sensor networks \cite{classif_liotta}, Internet of Things \cite{tnsm-classif-iot}, and automotive \cite{classif_pascale}.  
This notwithstanding, a great part of scientific literature, which faces the problem of data traffic classification through machine learning techniques, suffers from the datasets obsolescence issue. For many years, the only dataset available to the scientific community was the so-called KDD$99$, which is still broadly used today to validate ML-based algorithms and techniques when dealing with traffic classification problems. Unfortunately, this is an outdated  $20$-years-old dataset involving network attacks that have already been mitigated. An example is {\em satan probing}, an attack aimed at probing a computer for security loophole, based on the {\em satan} tool that was created at the end of the 1990’s, but is today no longer documented as a Web page.

Other examples include: {\em warezmaster/warezclient}, which exploits some old vulnerabilities of the anonymous FTP protocol, and {\em smurf attacks}, aimed at attaining default router settings that allowed directed broadcasts. Yet, currently, router vendors simply deactivate this functionality. 
An ameliorated version of KDD$99$ is known as NSL-KDD. However, despite providing some improvements (e.g. no redundant records, better balancing between training and test set), NSL-KDD is still not taking into account crucial information that characterizes novel cyber attacks. 

%%%% Gente che usa  KDD con varie tecniche ****
This notwithstanding, both KDD$99$ and NSL-KDD datasets are still broadly used to test some functionalities of NIDS systems that implement neural-based techniques. 
For instance, in \cite{precic_paper2,precic_paper3,ids_ann_19} the authors show the effectiveness of using an NIDS in conjunction with an artificial neural network (ANN) to improve the quality of traffic detection, where a validation stage onto the KDD$99$ dataset is performed. A deep learning approach is used in \cite{precic_paper1}, based on KDD$99$, to verify accuracy against an SVM methodology. A novel approach based on ANN (referred to as self-taught learning) is applied in \cite{precic_paper4} to enable an NIDS to detect previously unseen attacks via reconstructions made on unlabeled data. This work provides tests on both KDD$99$ and NSL-KDD. 
In \cite{precic_paper5} and \cite{precic_paper6} the authors adopt neural-based methods exploiting Self-Organizing Maps. In \cite{precic_paper7} and \cite{precic_paper8}, Learning Vector Quantization is coupled with SVM and k-Nearest Neighbor, respectively, to detect traffic anomalies. In \cite{ids_dnn_1,ids_dnn_2,ids_dnn3,ids_dnn4} deep neural networks concepts are applied to intrusion detection systems.

The KDD$99$ and NSL-KDD datasets have also been used to test a variety of non neural-based techniques such as Support Vector Machine \cite{kdd_svm1,kdd_svm2,kdd_svm3,kdd_svm4,kdd_svm5}, Principal Component Analysis \cite{kdd_pca1,kdd_pca2,kdd_pca3,kdd_pca4,kdd_pca5}, Decision Trees \cite{kdd_decision1,kdd_decision2,kdd_decision3,kdd_decision4,kdd_decision4,kdd_decision5}, and various unsupervised approaches \cite{kdd_unsup1,kdd_unsup2,kdd_unsup3,kdd_unsup4,kdd_unsup5,kdd_unsup6}.    

By contrast to the aforementioned works, some recent datasets are emerging from new testbeds which adhere more strictly to real-world network scenarios. For instance, the Cyber Range Lab of the Australian Centre for Cyber Security provides two recent datasets: the UNSW-NB15 dataset \cite{unsw2} which includes a mix of legitimate network activities and synthetic attacks, and the Bot-IoT dataset \cite{bot-iot1} which embeds normal and simulated network traffic gathered in an IoT-compliant testbed, along with various types of attacks.

Again, the datasets recently released by the Canadian Institute for Cybersecurity (CIC) \cite{cic} represent the state-of-the-art, in terms of both complexity and novelty of network attacks. These datasets have been created starting from an experimental testbed under controlled conditions \cite{icissp-dataset3}, whereby an attack network (including a router, a switch, and four PCs equipped with Kali Linux OS, which is a popular Linux distribution to perform penetration testing) is counterposed to a victim network (including routers, firewalls, switches, and various PCs equipped with Windows, Linux, and MacOS operating systems). An evolved version of this testbed has been designed to run on Amazon AWS \cite{aws}. In this case, the attacking infrastructure includes $50$ PCs, and the victim network includes $420$ PCs and $30$ servers. 
The implemented attacks encompass Distributed Denial of Service (DDoS), Portscan, Bruteforce, along with some typical Android-based network attacks injecting malicious codes such as adwares, malwares, and ransomwares.

The interest for such novel datasets is proven by novel works as detailed in the following.
In \cite{cic_paper1}, authors validate an artificial neural network system onto a CIC-released dataset to detect the malicious traffic generated by a botnet, out of the regular traffic. A hybrid neural network to detect anomalies in network traffic is proposed in \cite{cic_paper2}, where the CIC-IDS-2017 dataset has been exploited. Specifically focused on DDoS detection is the work in \cite{cic_paper3}, where a neural-based approach relying on the implementation of a simple Multi-Layer Perceptron is contrasted to the Random Forest technique. Again focused on DDoS detection is \cite{cic_paper4}, where some classic ML-based techniques (e.g. Na\"{\i}ve Bayes and Logistic Regression) are used to distinguish regular traffic from malicious one.    

\begin{table*}[h]
	\centering
	\caption{Prominent Related Work surveying ML techniques applied to Network Intrusion Detection.}
	\small
	\renewcommand{\arraystretch}{1.4}
	\begin{tabular}{|p{2.7cm}|p{3cm}|p{2.5cm}|p{6cm}|}
		\hline	
		
		\textbf{Authors} &  \textbf{Experiments} &  \textbf{Single/Multi Class} &  \textbf{Description} \\  \hline
		
		Nguyen et al. \cite{survey_pure1} & N/A & N/A & Classic survey on ML-based techniques with pointers to other works but with no experiments.  \\ \hline
		
		Boutaba et al. \cite{survey_pure2} & N/A & N/A & Survey on ML-based techniques applied to various networking-related problems (from traffic classification to routing or QoS/QoE management).  \\ \hline
		
		Hindy et al. \cite{survey_pure3} & N/A & N/A & Survey on IDS techniques taking into account ML algorithms such as ANN, K-means and SVM.  \\ \hline	
		
		Khraisat et al. \cite{survey_pure4} & N/A & N/A & Survey on Signature and Anomaly-based IDS techniques applying ML methods on NSL-KDD dataset. \\ \hline	
		
		Aldweesh et al. \cite{survey_pure5}  & N/A & N/A & Survey on Deep Learning techniques for IDSs with pointers to other works but with no experiments.  \\ \hline
		
		Fernandes et al. \cite{survey_pure6} & N/A & N/A & Survey on various techniques (ML, Statistical, Information Theory) for intrusion detection with pointers to other works but with no experiments.  \\ \hline
		
		Buczak et al. \cite{survey_pure7} & N/A & N/A & Survey on Data Mining and ML methods for intrusion detection with pointers to other works but with no experiments.  \\ \hline
		
		Tidjon et al. \cite{survey_pure8}  & N/A & N/A & Survey on various techniques (mainly ML-based) to be applied in intrusion detections with pointers to other works but with no experiments.  \\ \hline
		
		Azwar et al. \cite{azwar} & Performance analysis & Single Class & Non-homogeneous comparisons among various approaches (trees, NN) by using modern CIC-IDS$17$ dataset. \\ \hline
		
		Moustafa et al. \cite{moustafa_table}  & Performance analysis & Single Class & Holistic Survey on ML methods with experiments on feature reduction techniques (ARM, PCA, ICA).  \\ \hline
		
		Meena et al. \cite{Meena} & Time analysis & Single Class & Non-homogeneous comparisons between J48 and Na\"{\i}ve Bayes techniques on KDD and NSL-KDD datasets. \\ \hline
		
		Rama et al. \cite{kdd_rama} & Performance analysis, Time analysis (partial)  & Single Class & Non-homogeneous comparisons among various algorithms (e.g. J48,  Na\"{\i}ve Bayes, Bagging) on KDD and NSL-KDD datasets. \\ \hline
		
		Yin et al. \cite{kdd_rnn} & Performance analysis & Single/Multi Class & Non-homogeneous comparisons among various and different approaches (e.g. J48, ANN, SVM) on KDD$99$ dataset. \\ \hline		
		
		This work & Performance analysis, Time analysis & Single/Multi Class & Homogeneous comparisons among neural-based approaches (Deep, Weightless NN, LVQ, SOM) performed on modern CIC-IDS-2017/2018 datasets. \\ \hline
	\end{tabular}
				\label{tab:works}
\end{table*}

Going more precisely into the set of papers that share with the proposed work the purpose of surveying and/or comparing ML techniques for intrusion detection, we have gathered the prominent papers in Table \ref{tab:works}. The first column contains a pointer to the source; the second column reports the type of experimental analysis (if any); the third column highlights the type of datasets utilized (single/multi-class); the last column provides a brief description of the surveyed material, whereby the qualification ``non-homogeneous" refers to a comparison among techniques belonging to different families, thus, hardly comparable. 

Going beyond the adoption of novel datasets, we want to highlight the two most significant differences arising between the proposed review and the technical literature: First, to avoid dispersions, we prefer to focus on a specific family of ML techniques, namely the neural networks, so to guarantee more fair comparisons among outcomes. Such a focus allows us to better reveal the surprising behavior of weightless neural networks (traditionally exploited in the field of image classification) which exhibits an extraordinary advantageous accuracy/time complexity trade-off w.r.t. other NN-based methods. Then, we carry out an experimental analysis including performance and time complexity (this latter often neglected in technical literature) about all the described NN techniques; this effort overcomes the limit of gathering findings from various works (where different authors exploit different testbeds exhibiting different performance), thus, it results in a uniform vision of neural methods in the field of intrusion detection. Accordingly, Table \ref{tab:works} helps pinpointing such novel aspects of the present paper, as per last row.

\section{Neural-based techniques under scrutiny}
\label{sec:nn}

In this section we offer a brief recap of neural-based techniques that we then employ in Section V to perform our comparative assessment. We start with Multi-Layer Perceptron (MLP) which represents the basis for implementing ANNs and deep neural networks (DNN). Then, we consider WiSARD, one of the most representative algorithms of weightless neural networks. These typically provide interesting performance results, but have traditionally been applied in domains such as image classification rather than intrusion detection. We examine Learning Vector Quantization (LVQ) methods (with $3$ variants), where the notion of {\em codebook vector} will be introduced. Finally, we take into account the Self-Organizing Maps (SOM) technique.

\subsection{Multi-Layer Perceptron}

\begin{figure*}[t]
	\centering
	%	\captionsetup{justification=centering}
	\includegraphics[scale=0.65,angle=90]{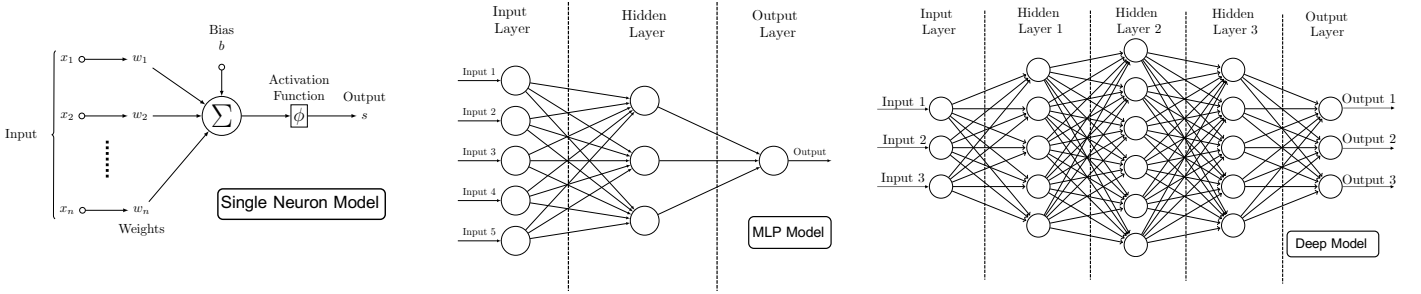}
	\caption{Single Neuron model (left panel). MLP model (middle panel). Deep NN model (right panel)}.
	\label{fig:nn}
\end{figure*}

The Multilayer Perceptron is one of the most representative type of feedforward-based ANNs, whereby there are no cycles arising in the connections among nodes. MLPs exhibit a fully connected structure of neurons, where each individual neuron calculates the weighted sum of $n$ inputs, adds a bias term $b$, and then applies an activation function $\phi(\cdot)$ to produce the output $s$, namely 
\beq
s=\phi \left( \sum_{i=1}^{n}w_i x_i +b \right).
\label{eq:neuron}
\eeq
Figure \ref{fig:nn} depicts \textit{i)} (left panel) the model of a single neuron (or single-unit perceptron) implementing (\ref{eq:neuron}), and \textit{ii)} (middle panel) a simplified structure of a typical MLP model  with $5$ neurons in the input layer, $3$ neurons in the hidden layer, and $1$ neuron in the output layer. In case of multiple hidden layers, MLP implements a Deep Neural Network (DNN) \cite{bengio} as reported in the right panel. 

The basic MLP functioning is described next. Input $I_n$ (see Fig.  \ref{fig:nn} - middle panel) activates the hidden layer (or layers) by following the forward activation direction, which is what justifies the term {\em feed-forwarding} neural network. Similarly, neurons in the hidden layers feed forward into output neurons, thus, an output value is obtained. It is useful to recall that the activation function is aimed at deciding whether or not a neuron would be activated, introducing some non-linearity into the neuron output. A variety of activation functions exist \cite{glorot} including: step function, linear, sigmoid, hyperbolic tangent, ReLU (Rectified Linear Unit), Softmax and many others. 

The MLP training stage is achieved through \textit{backpropagation}, a mechanism exploited to adjust weights of neurons aimed at progressively minimizing the error through Gradient Descent (GD), an iterative optimization algorithm useful to find local minima.  
Precisely, the purpose is to minimize an error function (e.g. least squares):
\beq
\mathcal{E}=\sum_{p \in \mathcal{P}} || t_p - s_p ||^2, 
\eeq
where $\mathcal{P}$ is the set of training patterns, $t_p$ is the target, and $s_p$ is the output for the example pattern $p$. 
The weight-updating rule, used to progressively compute the new weight $w_{new}$, is derived by evaluating the gradient $\partial \mathcal{E} / \partial w$, so that:
\beqa
w_{new}=w+\Delta w,  \nonumber \\ \nonumber \\
\Delta w = - \eta \frac{\partial \mathcal{E}}{\partial w} + \alpha \Delta w_{prev},
\eeqa
where: \textit{i)} $\eta$ is the {\em learning rate}, namely a hyperparameter lying in the range $(0,1)$ associated to the step size of the GD procedure (N.B.: too small $\eta$ implies difficulty of convergence, whereas too large $\eta$ could result in indefinite oscillations); \textit{ii)} $\alpha$ is defined as the {\em momentum}, a term lying in the range $(0,1)$ used to weight the influence of each iteration.
It is worth noting that, since derivative operations are involved in the backpropagation algorithm, non-linear activation functions (e.g. Sigmoid, ReLU) have to be exploited, whose derivative functions exist and are finite. 

%%%%%
In our MLP-based experiment we use two types of activation functions: ReLU for all the layers except for the output, and Softmax for the neurons in the output layer. ReLU has been proven to be one of the most effective activation functions when dealing with deep neural networks \cite{relu1,relu2} since it allows the whole network to converge rapidly. Conversely, Softmax is particularly suited for handling multiple classes in the output stage \cite{softmax}. 

%%%%

\subsection{Evolved Deep architectures}
Deep learning approaches allow to face a problem in a hierarchical way. Lower layers of the model are associated to a basic representation of the problem, whereas higher layers encode more complex aspects. Inputs feeding each layer of a DNN are manipulated through transformations which are parametrized by a number of weights. Although deep approaches are very promising, two main issues remain opened: first, training these architectures requires a significant computational power, and, then, the huge number of hyperparameters makes the tuning process very hard. In the following, we briefly discuss the most recent architectures relying on a deep-based approach.

	\textbf{Convolutional Neural Networks (CNNs)}: Such a technique \cite{cnn} takes inspiration from the human visual cortex, which embodies areas of cells responsive to particular regions of the visual field. This structure makes CNNs exceptionally suited for applications such as images classification or objects detection. Two main stages characterize the CNN lifecycle: feature extraction and classification. In the first stage, the so-called convolutional filters extract multiple features from the input and encode them in feature maps. The ``convolution'' is the mathematical operation consisting in an element-wise product and sum between two matrices, and has the main drawback of being hugely time consuming. The output of each convolutional layer feeds the activation function layer (e.g. ReLU) which produces an activation map by starting from a feature map. Finally, an optional pooling layer keeps only significant data. On the other hand, the classification stage is composed of a number of fully connected (or dense) layers followed by a Softmax output layer.

	\textbf{Recurrent Neural Networks (RNNs)}: It is a class of DNNs conceived on the basis of a work of Rumelhart \cite{rnn},  explicitly designed to deal with sequential data. Thus, RNNs are well suited for modelling language (intended as a sequence of interconnected words) in the field of the so-called natural language processing (NLP). 
	The key concept in RNNs is the presence of cycles, which represent the internal memory exploited to evaluate current data with respect to the past. Such a temporal dependency calls for the introduction of time-based hidden states obeying to:
	\beqa
	h(t)=\phi(h(t-1),x(t)),
	\eeqa 
	with the meaning that an internal state $h$ at time $t$ can be represented in terms of the input at time $t$ and the previous hidden state at time $t-1$. In this way, an RNN is helpful to predict the next element of a time series, or the next word in a sentence based on the number of the previous words. One of the main drawbacks of RNNs is dealing with long-term dependencies connected with transferring information from earlier to later times steps across too long sequences. To tackle this issue, two more sophisticated variants of RNNs have been introduced: LSTM and GRU.

	\textbf{Long Short-Term Memory (LSTM)}: It is a special RNN architecture \cite{lstm} conceived to learn long-term dependencies, namely, to store information for a long period of time. Basically, an LSTM unit (replacing an ordinary node in the hidden layer) is represented by a cell state in charge of carrying the information, and by three different \textit{gates} aimed at regulating the information flow: \textit{forget},  \textit{input}, and  \textit{output} gates. The \textit{forget} gate decides what information keep or discard on the basis of a forgetting coefficient calculated by input data $x(t)$ and the previous hidden state $h(t-1)$; the \textit{input} gate decides how to update the cell state; the \textit{output} gate decides which information has to be passed to the next unit, on the basis of input data and the previous hidden state. For the $i$-th LSTM unit, the hidden state at time $t$ can be expressed as:
	 \beqa
	 h^i(t)=out^i(t) \cdot tanh(c^i(t)) ,
	 \eeqa 
where $out^i(t)$ is an output gate which tunes the amount of memory, and $c^i(t)$ represents the cell state of LSTM unit $i$ at time $t$.

	\textbf{Gated Recurrent Unit (GRU)}: It is a lightweight version of LSTM \cite{gru} and has only two gates: \textit{update} gate and \textit{reset} gate. The former plays a similar role of forget and input gates of LSTM; the latter is exploited to decide how many past data to forget. In the GRU model, the hidden state $h^i(t)$, corresponding to the $i$-th GRU unit, can be expressed as a linear interpolation between the hidden state at time $t-1$ and the candidate activation (or hidden state) $\widetilde{h^i}(t)$ viz.
	 \beqa
	 h^i(t)=(1-z^i(t)) h^i(t-1) + z^i(t)  \widetilde{h^i}(t),
	 \eeqa 
where $z^i(t)$ is the update gate which decides about the updating amount of its candidate activation.

\begin{figure}[t]
	\centering
	%	\captionsetup{justification=centering}
	\includegraphics[scale=0.36,angle=90]{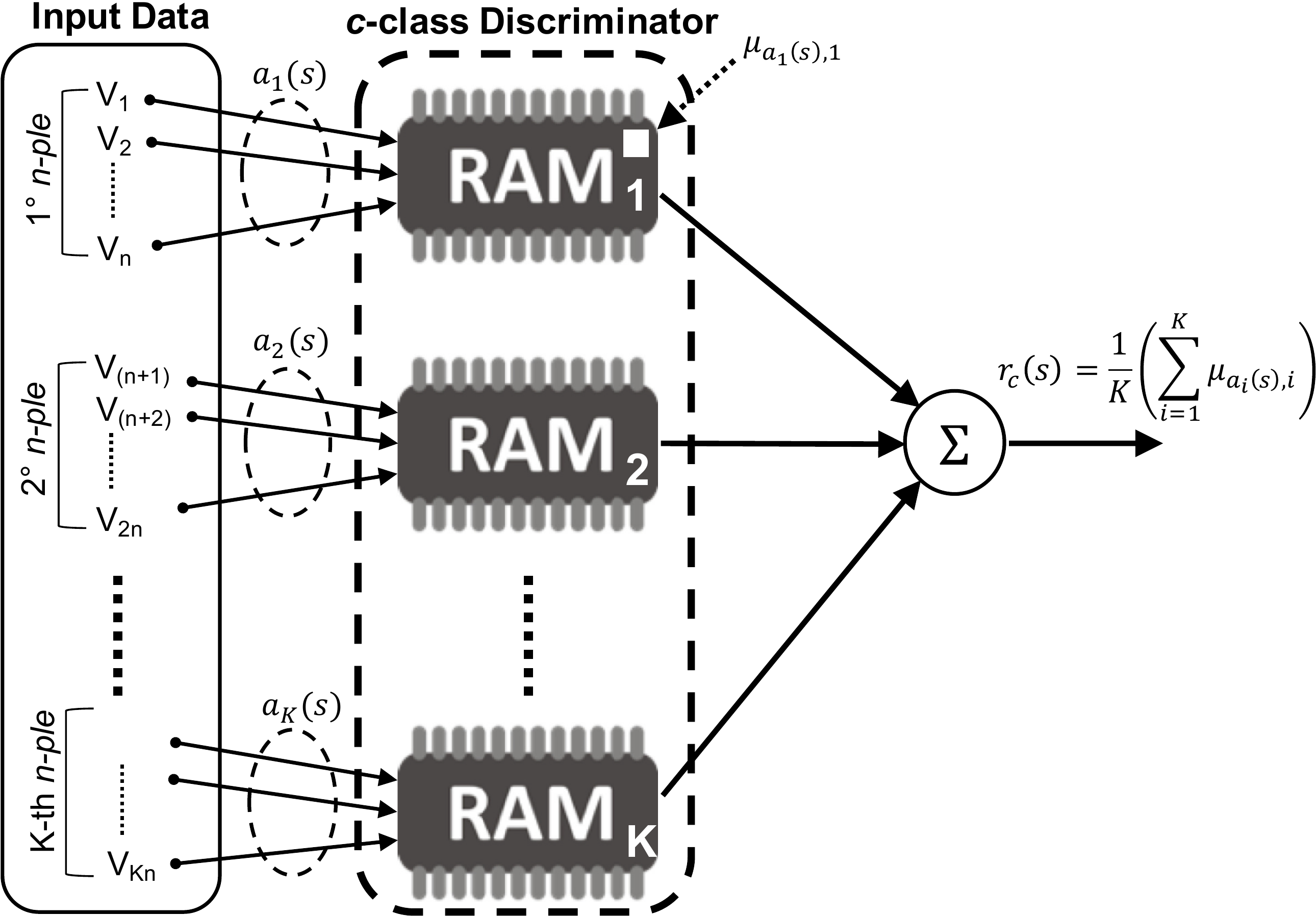}
	\caption{Model of a single class discriminator for the WiSARD algorithm.}
	\label{fig:discri}
\end{figure} 

\subsection{WiSARD}

WiSARD\footnote{\textbf{Wi}lkes, \textbf{S}tonham and \textbf{A}leksander \textbf{R}ecognition \textbf{D}evice } is a supervised method \cite{wisard,wisard_new} that was originally conceived for image classification, but has recently been proven to be effective in more general multi-class problems. 

WiSARD falls under the class of weightless neural networks (WNNs) since it exploits lookup tables, instead of weights, to store the functions evaluated by the individual neurons. WNNs rely on a mechanism inspired to the random access memory (RAM) encoding functionalities, since input data are transformed into binary. This process has a noteworthy benefit in terms of time complexity due to the use of Boolean logic, which can be further improved by exploiting pipelining and parallelism.  
In short, WiSARD is composed of a set of classifiers (or {\em discriminators}), each one in charge of learning binary patterns associated to a particular class. In turn, a discriminator is composed of a set of neurons, referred to as RAM neurons as depicted in Fig. \ref{fig:discri}. Similarly to conventional RAM circuits, a RAM neuron (a.k.a. $n$-tuple neuron) can be interpreted as a RAM having $2^n$ memory locations addressed by $n$ address lines (inputs) representative of neuron connectivity. 

During the training stage, each RAM neuron learns the occurrences of $n$-tuple vectors extracted from a training pattern, and stores them in a memory cell (equivalent to a RAM writing operation). Precisely, given $\mu_{a,i}$ the memory cell with address $a$ in the $i$-th RAM (initially empty), the following update rule holds:
\beq
\mu_{a,i}=\theta  \left(   \sum_{{p}\in{\mathcal{P}}} \delta_{a,a_i (p)}   \right),
\eeq
where: function $\theta(z)$ amounts to $z$ if $0\le z \le 1$ and to $1$ if $z>1$; $p$ is a pattern defined in the training set $\mathcal{P}$; $a_i (p)$ is the address generated starting from pattern $p$; $\delta$ is the Kronecker delta function, amounting to $1$ if $a=a_i (p)$ and 0 elsewhere.  

The classification stage (equivalent to a RAM reading operation) consists of classifying an unseen pattern $s$ by assigning $s$ to a class $c$, provided that the correspondent discriminator exhibits the highest output, namely
\beq
\argmax_{c} \left( \sum_{i=1}^{K} \mu_{a_i(s),i} \right),
\eeq
whereas, the response of the $c$-th class discriminator on pattern $s$ is:
\beq
r_c(s)=\frac{1}{K} \sum_{i=1}^{K} \mu_{a_i(s),i}.
\eeq
Accordingly, since WiSARD is made of a set of discriminators, the overall response of a set of discriminators trained on $N$ classes produces a response vector $\bold{r}(s)=[r_{c_1}(s),\dots,r_{c_N}(s)]$.

\subsection{Learning Vector Quantization}

Learning Vector Quantization (LVQ) \cite{lvq} directly stems from classical Vector Quantization, a signal-approximation method aimed at representing the input data vectors $x  \in \mathbb{R}^n$ through a finite number of {\em codebook} vectors $m_i  \in \mathbb{R}^n, i=1,2,\dots,k$. The goal is to find the codebook vector ${m_v}$ that best approximates $x$, usually in terms of Euclidean distance, namely:
\beq
v= \argmin_{i} || x - m_i ||. 
\label{eq:lvq}
\eeq
The main purpose of LVQ is to define {\em class regions} (over the input data space), each one containing a subset of a similarly labeled codebook. Accordingly, it is possible to pinpoint hyperplanes between neighboring codebook vectors defining the so-called {\em quantization regions}.  
By assuming that all samples of ${x}$ derive from a finite set of classes $S_k$, the idea is to \textit{i)} assign a subset of codebook vectors to each class $S_k$, and \textit{ii)} search for ${m_v}$ having the smallest Euclidean distance from ${x}$.  
Different versions of LVQs have been introduced in the literature with slight differences, as introduced in the following.
\vspace{6pt}

\subsubsection{LVQ1}

Assume that $x(t)$ is an input sample and $m_i(t)$ contains sequential values of $m_i$ ($t=0,1,2,\dots$). The LVQ1 algorithm allows to find values of $m_i$ in (\ref{eq:lvq}) that asymptotically minimize errors, and is defined by the following set of equations:
\beqa
	&m_v(t+1)&=m_v(t)+\alpha(t)\left[x(t)-m_v(t)\right], \label{eq:lvq_1}  \\
	&m_i(t+1)&=m_i(t), \label{eq:lvq_2} 
	\label{eq:lvq1}
\eeqa
where (\ref{eq:lvq_1}) refers both for cases that $x$ and $m_v$ belong to the same or different classes, (\ref{eq:lvq_2}) holds for $ i \neq v$, and $\alpha(t)$ is the learning rate. This algorithm admits also an optimized version (a.k.a. OLVQ1), where an individual learning rate $\alpha_i$ is assigned to $m_i$, thus, basic $m_v(t+1)$ equation in (\ref{eq:lvq1}) becomes:
\beq
m_v(t+1)=\left[ 1- c(t)\alpha_v(t) \right] m_v(t) + c(t)\alpha_v(t)x(t), 
\eeq 
where $c(t)=+1 [-1]$ if the classification is correct [wrong]. 
\vspace{6pt}
\subsubsection{LVQ2}

Differently from the standard procedure implemented in LVQ1, here two codebook vectors $m_i$ and $m_j$ (belonging to the correct and to the wrong class, respectively) are simultaneously updated. In this case, $x$ must fall within a ``window" defined around the hyperplanes of $m_i$ and $m_j$. The correspondent algorithm is defined by the following set of equations:

\beqa
	&m_i(t+1)&=m_i(t)-\alpha(t)\left[ x(t) -m_i(t) \right], \label{eq:lvq2_1} \\
	&m_j(t+1)&=m_j(t)-\alpha(t)\left[ x(t) -m_j(t) \right], \label{eq:lvq2_2}
\eeqa
where (\ref{eq:lvq2_1}) holds when $x$ and $m_i$ belong to the same class, (\ref{eq:lvq2_2}) holds when $x$ and $m_i$ belong to different classes, and with $m_i$ and $m_j$ being the two closest codebook vectors to $x$. Moreover, $x$ has to fall within a window of relative width $w$ if 
\beq
\textnormal{min} \left(  \frac{e_i}{e_j}, \frac{e_j}{e_i}  \right) > k, 
\eeq
where $e_i$ and $e_j$ represent, respectively, the Euclidean distances of $x$ from $m_i$ and $m_j$, and $k=\frac{1-w}{1+w}$.
\vspace{6pt}
\subsubsection{LVQ3}
\label{sec:lvq3}
This variant of LVQ2 admits the same set of equations (\ref{eq:lvq2_1}), (\ref{eq:lvq2_2}), with the difference that the learning rate is weighted with a parameter $\epsilon$, whose best values are found empirically to lie in the [0.1-0.5] interval of values:
\beq
m_h(t+1)=m_h(t)+\epsilon \alpha(t) \left[x(t)-m_h(t)\right], 
\eeq  
for $k \in {i,j}$ if $x$, $m_i$, and $m_j$ belong to the same class.

\subsection{Self-Organizing Maps}
SOM takes inspiration from a particular adaptive characteristic that allows the human brain to empower the experience. Specifically, given a physical stimulus (namely, the input) which activates multiple neurons of a certain brain area in parallel, those neurons that are more sensitive to the input stimulus will go about influencing all other neighboring neurons. 

This biological mechanism has led to designing SOM as a nonlinear mapping of high-dimensional input data onto elements of a low-dimensional array (a.k.a. \textit{lattice}) \cite{kohonen}, according to a principle known as competitive learning.  
%%%%%%% Competitive Leraning Concepts%%%%%
This is different from the classic ANN-based approach, where weights are updated to iteratively minimize errors. 
In competitive learning, several neurons are fed with the same input (in parallel) and compete to become the possible ``winner" in relation to that particular input.

According to this strategy, and assuming that the weight vector of neurons has the same dimensionality as the input, the output neuron activation increases with larger similarity between the weight vector of the neuron and the input \cite{aggarwal}. Precisely, by considering a network composed of $k$ neurons (with $k<<n$ being $n$ the data set size), an input vector $x=\left[ x_1, x_2, \dots, x_n \right]^T \in \mathbb{R}^n$ and a reference (or weight) vector $m_i=\left[m_{i1}, m_{i2}, \dots, m_{in} \right]^T \in \mathbb{R}^n$ associated with neuron $i$, the competitive approach can be summarized according to the following steps:
\begin{itemize}
	\item[] \textit{i)} Compute the smallest Euclidean distance $d=\argmin_{i}$ $\lvert\lvert$ $x$ $-$ m$_i$ $\lvert\lvert$  $\forall i \in (1,\dots,m)$ having the meaning of the activation value for the $i$-th neuron. The neuron having $d$ is declared as winner.
	\item[] \textit{ii)} The $i$-th neuron is updated according to the following rule: $m_i(t+1)=m_i(t) h(t)\left[ x(t) - m_i(t) \right]$ where $t=(0,1,\dots)$ is an integer discrete time reference, whereas $h(t)$ is the {\em neighborhood function} defined over the lattice points that is typically implemented through a Gaussian kernel (the same as then one exploited for the present experimental analysis):
	\beq
	h(t)=\alpha(t)\cdot \textnormal{exp} \left( - \frac{\lvert\lvert \ell_c - \ell_i \lvert\lvert^2}{2 \sigma^2 (t)}  \right),
	\eeq
	where: $\alpha(t)$ is the learning rate factor, $\ell_c \in \mathbb{R}^2$ and $\ell_i \in \mathbb{R}^2$ are, respectively, the location vectors of neurons $c$ and $i$ mapped on the lattice structure (supposed to be bi-dimensional), and $\sigma(t)$ represents the kernel width. 
\end{itemize}
Although SOM may be considered as the unsupervised counterpart version of LVQ,
a common way to exploit SOM for classification purposes is to consider a supervised version (sometimes dubbed as LVQ-SOM \cite{kohonen} and implemented in our analysis) relying on the following consideration: once we know that each training sample $x(t)$ and $m_i(t)$ have been assigned to specific classes, $h(t)$ has to be set to positive if $x(t)$ and $m_i(t)$ belong to the same class, and negative if they belong to different classes, by taking into account that such a rule is applied for each $m_i(t)$ in the neighborhood of the winner.

\begin{table}[htp]
	\centering
	\caption{Synthetic description of adopted features}
	\resizebox{.45\textwidth}{!}{
		\begin{tabular}{c|l}
			\textbf{Family} & \multicolumn{1}{c}{\textbf{List of features}} \\
			\midrule
			\multirow{5}[2]{*}{\textbf{Coarse-Grained}} 
			& 1. Source IP Address \\
			& 2. Destination IP Address \\
			&  3. Source Port \\
			& 4. Destination Port \\
			&  5. Transport Protocol Type \\
			\midrule
			\multirow{23}[2]{*}{\textbf{Time-Based}} 
			& 6. Flow duration \\
			&  7. Average inter-arrival times (IAT) between two flows \\
			&  8. IAT standard deviation (std) between two flows \\
			&  9. IAT max between two flows \\
			&  10. IAT min between two flows \\
			&  11. IAT tot between two pkts sent in fwd direction \\
			&  12. IAT avg between two pkts sent in fwd direction \\
			&  13. IAT std between two pkts sent in fwd direction \\
			&  14. IAT max between two pkts sent in fwd direction \\
			&  15. IAT min between two pkts sent in fwd direction \\
			&  16. IAT tot between two pkts sent in bwd direction \\
			&  17. IAT avg between two pkts sent in bwd direction \\
			&  18. IAT std between two pkts sent in bwd direction \\
			&  19. IAT max between two pkts sent in bwd direction \\
			&  20. IAT min between two pkts sent in bwd direction \\
			&  21. Avg time a flow was active before becoming idle \\
			&  22. Std time a flow was active before becoming idle \\
			&  23. Min time a flow was active before becoming idle \\
			&  24. Max time a flow was active before becoming idle \\
			&  25. Avg time a flow was idle before becoming active \\
			&  26. Std time a flow was idle before becoming active \\
			&  27. Min time a flow was idle before becoming active \\
			&  28. Max time a flow was idle before becoming active \\
			\midrule
			\multirow{6}[2]{*}{\textbf{Flow-Based}} 
			&  29. Flow byte rate  \\
			&  30. Flow pkt rate  \\
			&  31. Avg no. of pkts in a sub-flow in fwd direction    \\
			&  32. Avg no. of pkts in a sub-flow in bwd direction    \\
			&  33. Avg no. of bytes in a sub-flow in fwd direction    \\
			&  34. Avg no. of bytes in a sub-flow in bwd direction    \\
			\midrule
			\multirow{20}[2]{*}{\textbf{Packet-Based}} 
			& 35. Tot pkts in the fwd direction \\
			& 36. Tot length of pkts in the fwd direction \\
			& 37. Avg length of pkts  \\
			& 38.  Std length of pkts  \\
			&  39. Variance length of pkts  \\
			&  40. Avg length of pkts in the fwd direction \\
			&  41. Std length of pkts in the fwd direction \\
			&  42. Max length of pkts in the fwd direction \\
			&  43. Min length of pkts in the fwd direction \\
			&  44. Tot pkts in the bwd direction \\
			&  45. Tot length of pkts in the bwd direction \\
			&  46. Avg length of pkts in the bwd direction \\
			&  47. Std length of pkts in the bwd direction \\
			&  48. Max length of pkts in the bwd direction \\
			&  49. Min length of pkts in the bwd direction \\
			&  50. Avg no. of pkt bulk rate in fwd direction    \\
			&  51. Avg no. of pkt bulk rate in bwd direction    \\
			&  52. Fwd pkt rate   \\
			&  53. Bwd pkt rate   \\
			&  54. Min segment size in fwd direction \\
			\midrule
			\multirow{6}[2]{*}{\textbf{Byte-Based}} 
			&  55. Avg no. of byte rate in fwd direction    \\
			&  56. Avg no. of byte rate in bwd direction    \\
			&  57. No. of bytes sent in init win in fwd direction \\
			&  58. No. of bytes sent in init win in bwd direction \\
			&  59. Tot bytes used for headers in fwd direction \\
			&  60. Tot bytes used for headers in bwd direction \\
			\midrule
			\multirow{19}[2]{*}{\textbf{Flag-Based}} 
			& 61. No. of times URG flag set in fwd direction \\
			& 62. No. of times PSH flag set in fwd direction \\
			& 63. No. of times FIN flag set in fwd direction \\
			&  64. No. of times SYN flag set in fwd direction \\
			&  65. No. of times RST flag set in fwd direction \\
			&  66. No. of times ACK flag set in fwd direction \\
			&  67. No. of times URG flag set in bwd direction \\
			&  68. No. of times PSH flag set in bwd direction \\
			&  69. No. of times FIN flag set in bwd direction \\
			&  70. No. of times SYN flag set in bwd direction \\
			&  71. No. of times RST flag set in bwd direction \\
			&  72. No. of times ACK flag set in bwd direction \\
			&  73. PSH flag count \\
			&  74. FIN flag count \\
			&  75. SYN flag count \\
			&  76. RST flag count \\
			&  77. ACK flag count \\
			&  78. ECE flag count \\
			%&  CWE flag count \\
			\bottomrule
		\end{tabular}%
	}
	\label{tab:superfeatures}%
\end{table}%

\begin{table*}[t]
	\centering
	\caption{Optimized hyperparameters for the exploited algorithms.}
	\small
	\renewcommand{\arraystretch}{1.35}
	\begin{tabular}{|p{2.1cm}|p{8cm}|}
		\hline	
		\textbf{Algorithm} &  \textbf{Optimized hyperparameters and models info}  \\  \hline
		MLP-$1$, Deep-$2$, Deep-$3$ & $\cdot$  Stochastic Gradient Descent with adaptive hyperparams. (Adam version - \cite{adam})    \\ 
		& $\cdot$ LR (learn. rate)=$0.001$  \\ 
		& $\cdot$ Number of weights=$2000$  \\ 
		& $\cdot$  Exp. Decay (first moment estimate)=$0.9$  \\ 
		& $\cdot$  ReLU activation function \\
		& $\cdot$ Neurons per hidden layer: MLP-$1$($26$); Deep-$2$($23$,$10$); Deep-$3$($20$,$16$,$11$) \\  \hline		
		Convolutional & $\cdot$ $7$ filters $(4x1)$, $8$ neurons fully connected \\
		& $\cdot$ $1$ Pooling layer with Pooling size = 2 \\
		& $\cdot$ $1$ Dropout layer with Dropout rate = 0.3 \\
		& $\cdot$ ReLU activation function \\ \hline
		Recurrent-type & $\cdot$ $18$ Recurrent units (RNN), $10$ neurons fully connected \\
		& $\cdot$ $6$ LSTM units, $8$ neurons fully connected\\
		& $\cdot$ $8$ GRU units, $10$ neurons fully connected \\ \hline
		WiSARD & $\cdot$ Batch Size=$100$ \\
		& $\cdot$ Resolution (in bit) per neuron: $8$ \\ \hline
		LVQ(1,2,3) & $\cdot$ Batch Size=$100$ \\ 
		& $\cdot$ LR=$0.3$ 	\\ 
		& $\cdot$ Codebook Vectors=$20$ \\ 
		& $\cdot$ Window Size (for LVQ$2$ and LVQ$3$)=$0.3$ \\ 
		& $\cdot$ $\epsilon$ (for LVQ$3$)=$0.1$ \\ 
		\hline												   
		SOM 	                                 & $\cdot$ Batch Size=$100$ \\
		& $\cdot$ LR=$0.3$ 	\\ 
		& $\cdot$ Hexagonal Topology with Neighborhood Size = $8$ \\ 
		& $\cdot$ Neighborhood Function: Gaussian \\ \hline
	\end{tabular}
	\label{tab:hyperparams}
\end{table*}

\section{The experimental Datasets}
\label{sec:dataset}

In this section we present further details about datasets employed in our experimental analysis. As already remarked in Section \ref{sec:rw}, the datasets must convey the most recent information about a variety of cyber attacks across data networks. We adopted datasets released from CIC \cite{cic}, and precisely: the {\em DDoS } dataset, containing traffic relating to distributed denial of service attacks designed to saturate network resources; the {\em Portscan} dataset, including attempts of Portscan, a technique used to discover open ports on network devices; the {\em WebAttack} dataset which encompasses various malicious traffic ranging from Cross-Site Scripting to Sql Injection; the {\em TOR} dataset, a collection of network traffic traversing the anonymous TOR circuit often conveying malicious information; and the {\em Android} dataset, embedding a number of mobile (Android-based) adwares. 

These datasets have been used to perform two kinds of neural-based analyses: a {\em single-class} analysis, aimed at classifying the network traffic as benign or malign (binary information); and a {\em multi-class} analysis, aimed at distinguishing more classes of attacks. Each dataset has been cleaned and re-balanced through a Python routine designed from scratch, and contains $2\cdot10^4$ instances and $78$ features, which are then grouped in $6$ macro-classes, as indicated in the following (refer to the Table \ref{tab:superfeatures} for an exhaustive list of features):

\begin{itemize}
	\item \textbf{Coarse-grained features:} Source and Destination Port, Protocol Type, Source and Destination IP Address; 
	\item \textbf{Time-based features:} Backward/Forward inter-arrival times between two flows, duration of active flow (mean, std, min, max), duration of an idle flow (mean, std, min, max), etc.; 
	\item \textbf{Flow-based features:} Length of a flow (mean, min, max, etc.); 
	\item \textbf{Packet-based features:} Backward/Forward number of packets in a flow, Backward/Forward length of packets in a flow (mean, std, min, max, etc.);   
	\item \textbf{Byte-based features:} Backward/Forward number of bytes in a flow, Backward/Forward number of bytes used for headers, etc.;  
	\item \textbf{Flag-based features:} Number of packets with active TCP/IP flags (SYN, FIN, PUSH, RST, URG, etc.). Please note that, for example, feature $62$ ($68$) indicates the no. of times the PSH flag is set in the forward (backward) direction, whereas feature $73$ indicates the overall number of packets containing such a flag.
\end{itemize} 
%\vspace{5pt}
These features allow to derive statistical information that cannot be hidden in a possible malicious flow. Let us consider, for instance, a DDoS attack. This is typically  designed to overwhelm the resources of a target network \cite{ddos1bis,ddos2} by conveying legitimate information (e.g. trivial HTTP requests in case of an application-layer DDoS attack) which would pass unnoticed to a classic signature-based detection systems. However, DDoS attacks are designed to be a coordinated effort where multiple, malicious entities (a.k.a. {\em bots}) send few, tiny packets to the target. 
%Typically, these follow specific patterns, which can be picked up by ML engines. 
Similarly, the structure of a Portscan attack, characterized by a very quick scan of the victim's destination port by the network attacker, can be learned by crossing information of destination port and time-based information.    

\begin{figure*}[t!] 
	\centering
	\begin{tabular}{cc}
		\subfloat[]{\includegraphics[scale=0.51]{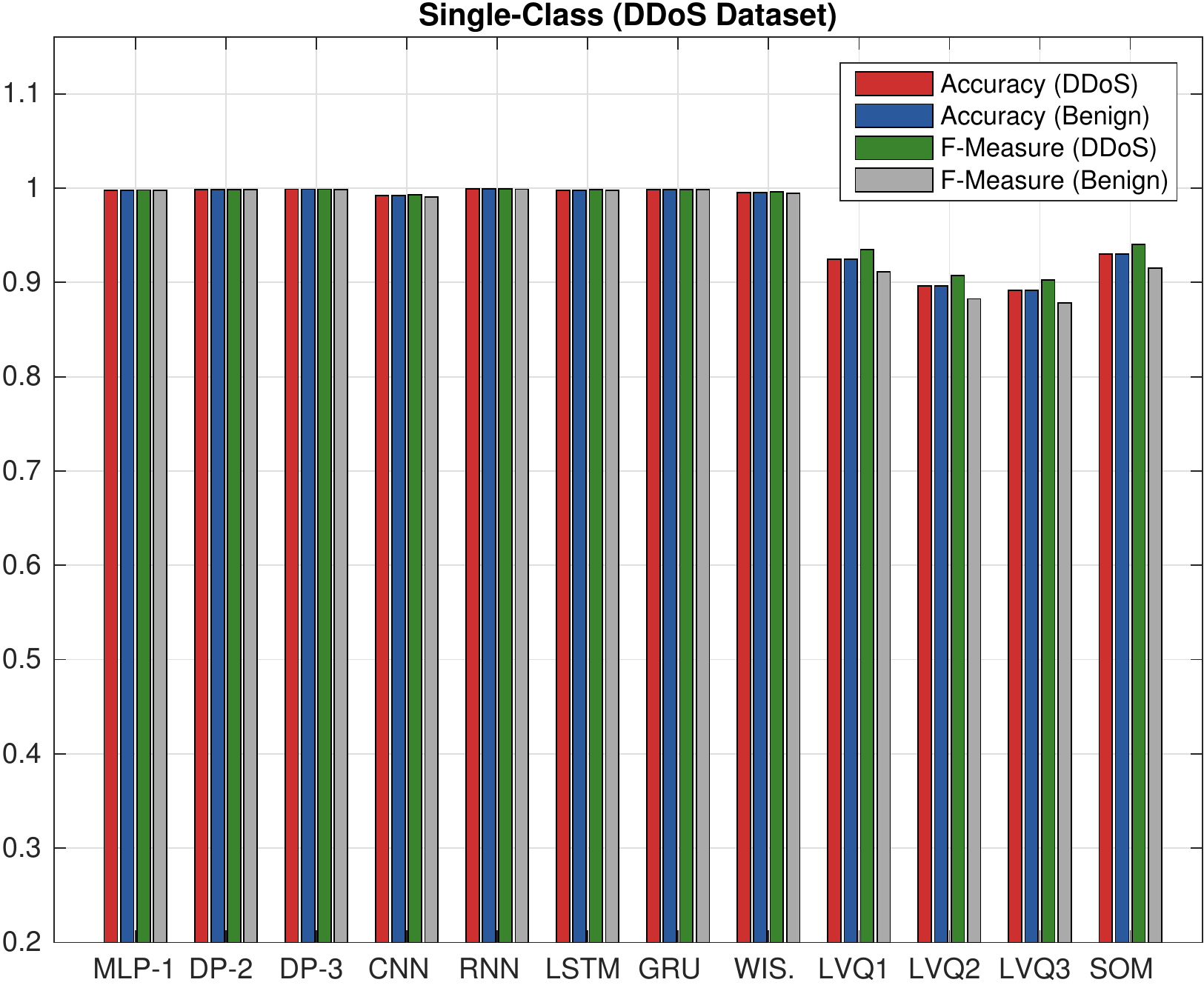}}  \hspace{2.5mm}
		\subfloat[]{\includegraphics[scale=0.51]{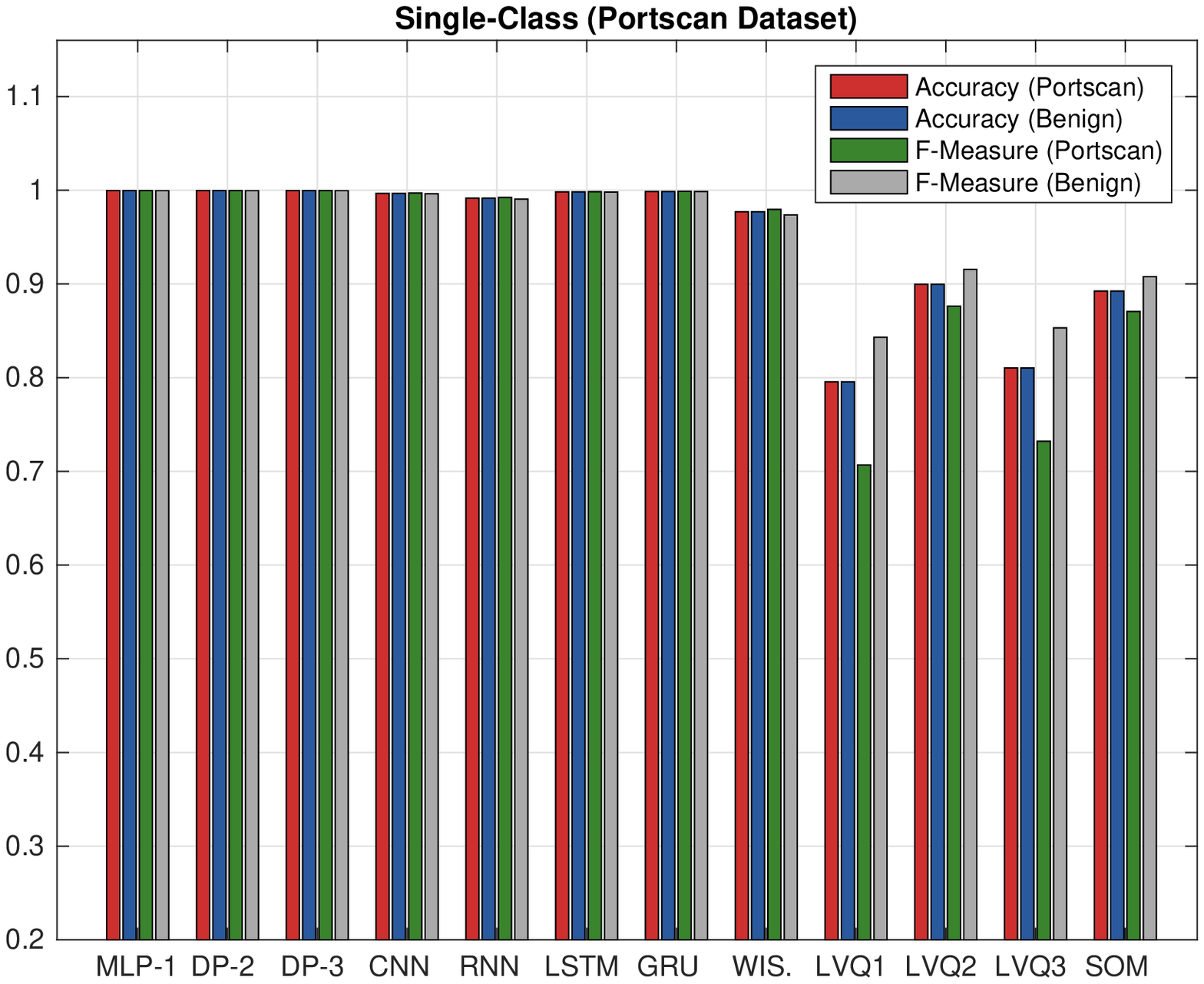}}  \hspace{2.5mm} \\
		\subfloat[]{\includegraphics[scale=0.51]{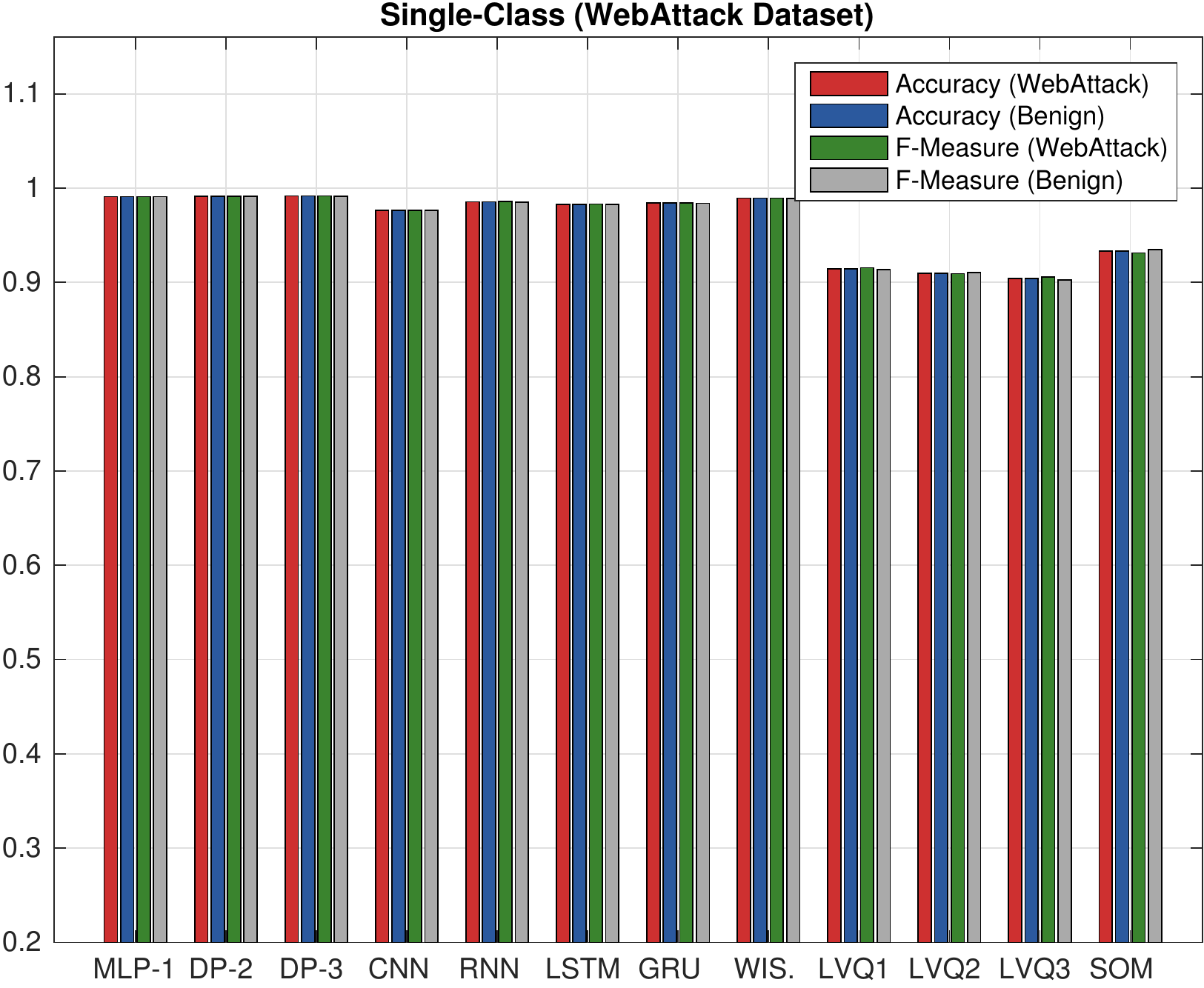}}  \hspace{2.5mm}
		\subfloat[]{\includegraphics[scale=0.51]{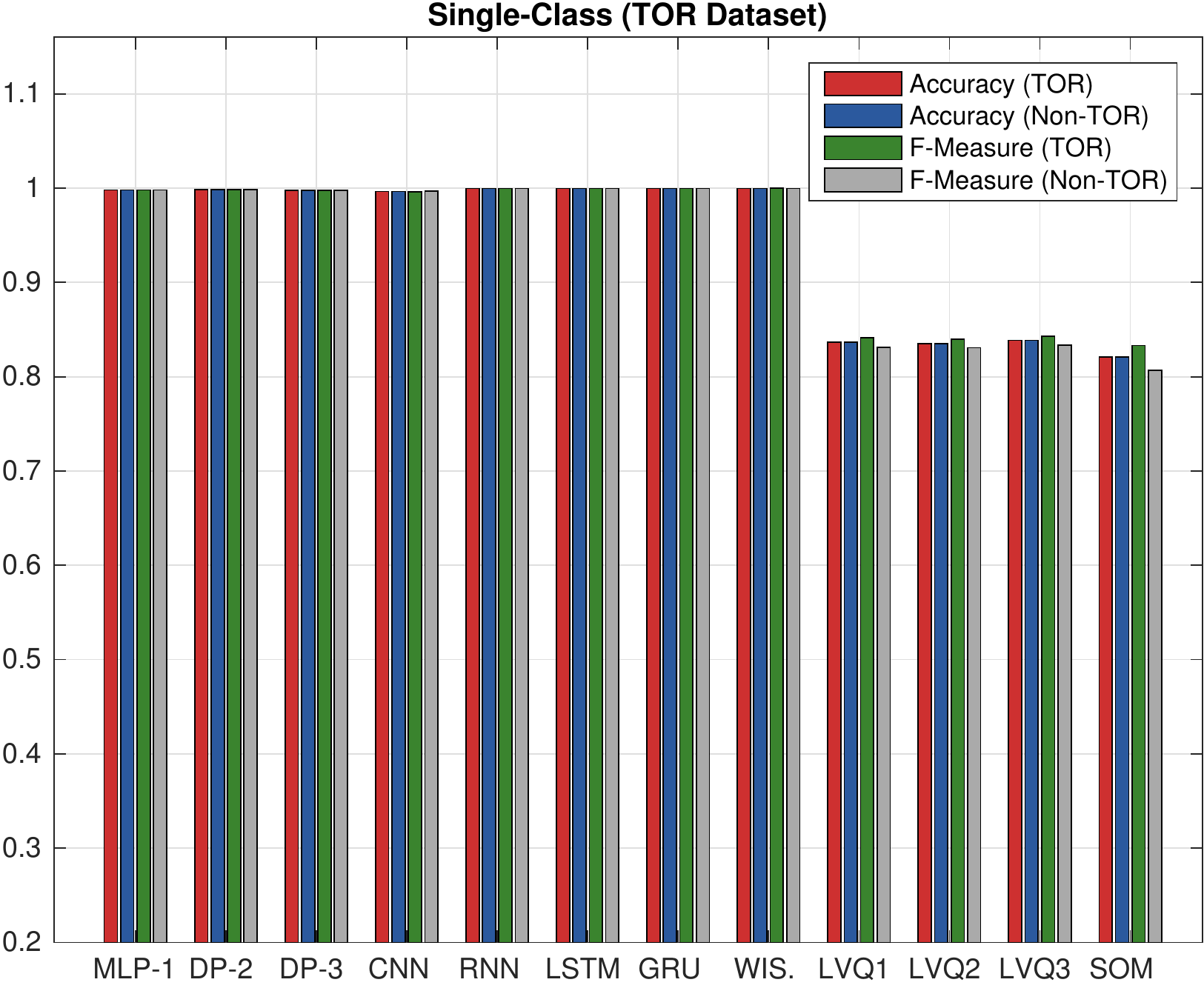}}  \hspace{2.5mm} 
	\end{tabular}
	\caption{Performance in terms of Accuracy/F-Measure for different single-class datasets: (a) DDoS; (b) Portscan, (c) WebAttack, (d) TOR.}
	\label{fig:perf_single}
\end{figure*}

% Table generated by Excel2LaTeX from sheet 'test'

\section{Experimental-based Assessment}
\label{sec:experim} 

The main purpose of our experimental analysis is to compare the neural techniques introduced in Sect. \ref{sec:nn} across the datasets described in Section IV. 
Aimed at a fair comparison, we adopt a $10$-fold cross validation for each experiment. The model structure for each algorithm and the pertinent hyperparameters are summarized in Table \ref{tab:hyperparams}.

\noindent The whole assessment comprises:
\begin{itemize}
	\item \textbf {Performance analysis}: obtained by evaluating two metrics typically used in the field of traffic classification \cite{accuracy2,tnsm-perf}, viz.
	\SubItem {\em Accuracy}: ratio of correctly predicted observations to the total, calculated by
	 	\beq
	 \frac{TP+TN}{TP+TN+FP+FN};
	 	\eeq
	\SubItem {\em F-Measure}: an indicator of a per-class performance, calculated by
	\beq
	2 \cdot \frac{\textnormal{Precision} \cdot \textnormal{Recall} }{\textnormal{Precision}+\textnormal{Recall}},
	\eeq
	where Precision is the ratio of correctly classified traffic over the total predicted traffic in a class, and Recall is the ratio of correctly classified traffic over all ground truth traffic in a class. 
	\item \textbf {Time complexity analysis}: derived by measuring the whole classification process (including training time) as the number of instances grows from $10^3$ to $2\cdot10^4$.
\end{itemize}

We perform the overall analysis on a PC equipped with Intel CoreTM i5-7200U CPU@ 2.50GHz CPU and 16 GB of RAM. 
For the sake of convenience, we split our analysis in two: the single-class and the multi-class cases. 

\begin{figure*}[t!] 
	\centering
	\begin{tabular}{cc}
		\subfloat[]{\includegraphics[scale=0.47]{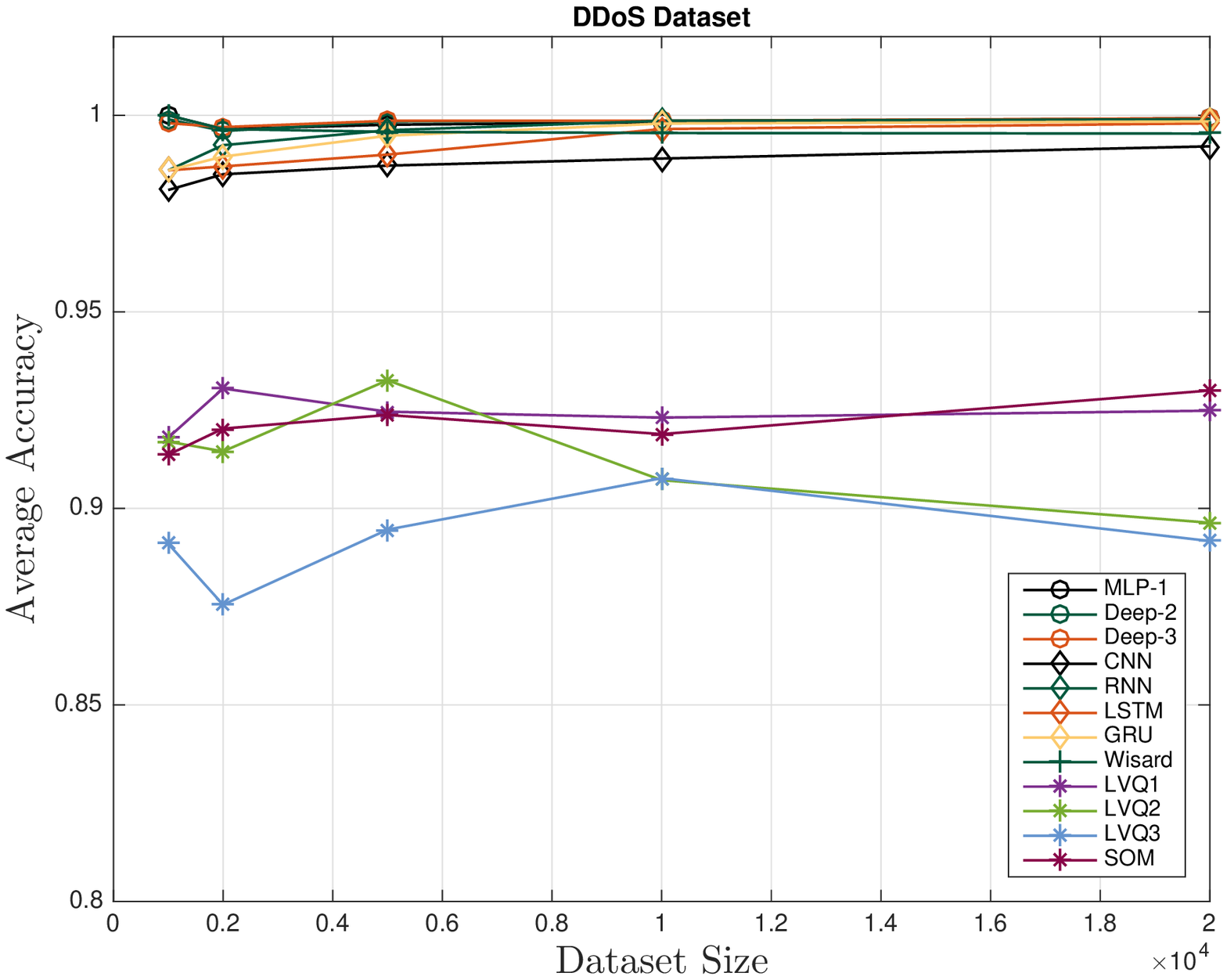}}  \hspace{2.5mm}
		\subfloat[]{\includegraphics[scale=0.47]{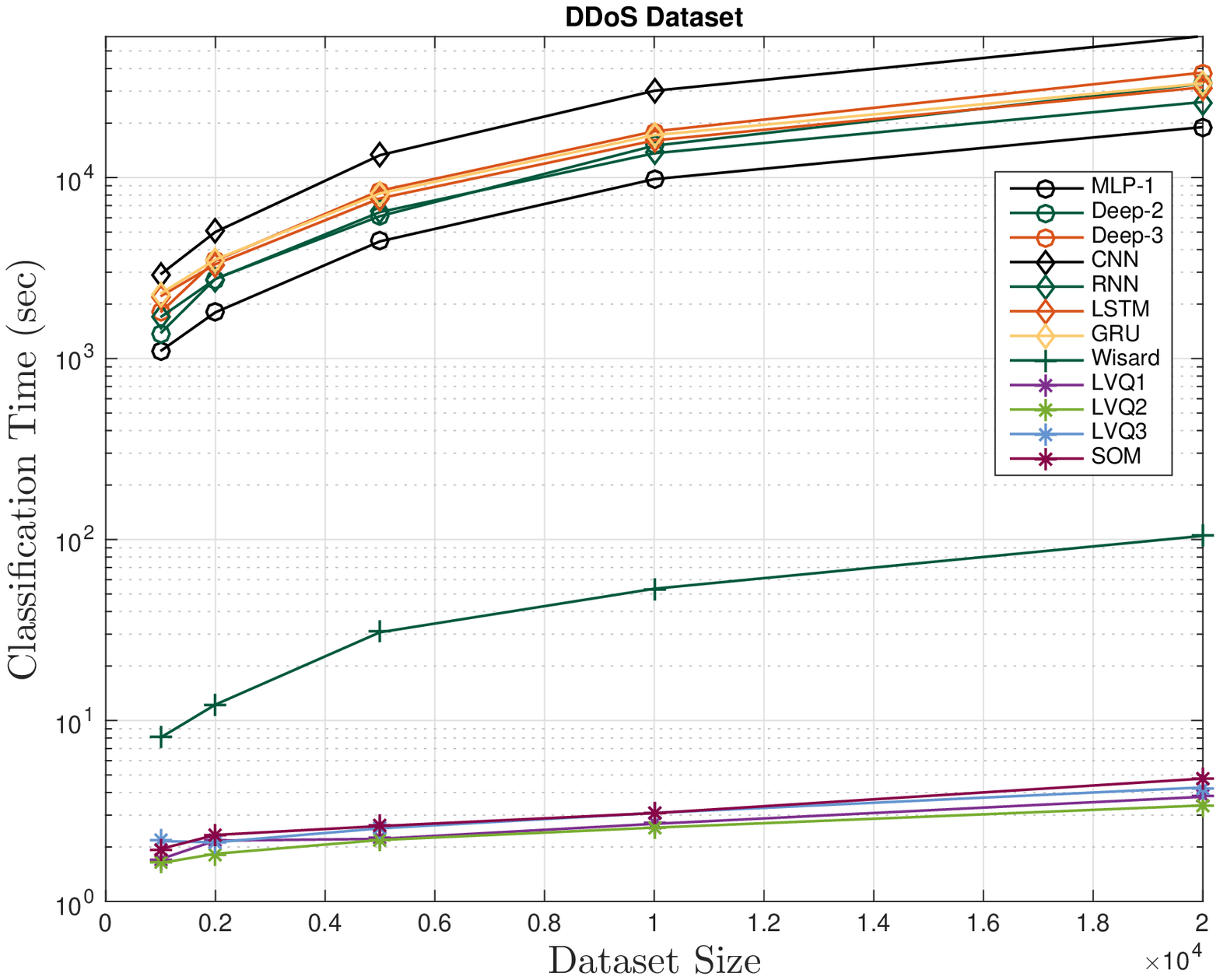}}   \hspace{2.5mm} \\
		\subfloat[]{\includegraphics[scale=0.47]{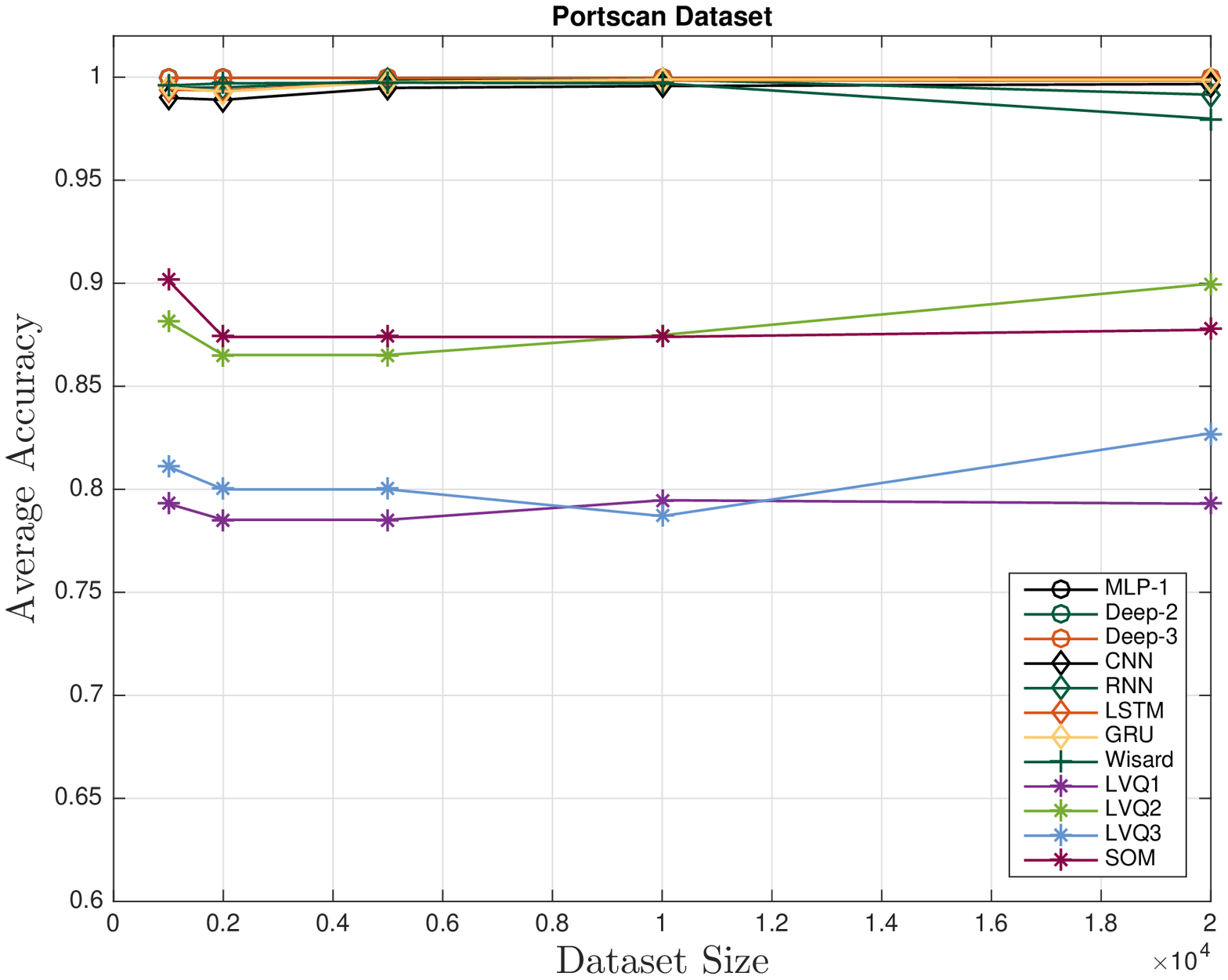}}  \hspace{2.5mm}
		\subfloat[]{\includegraphics[scale=0.47]{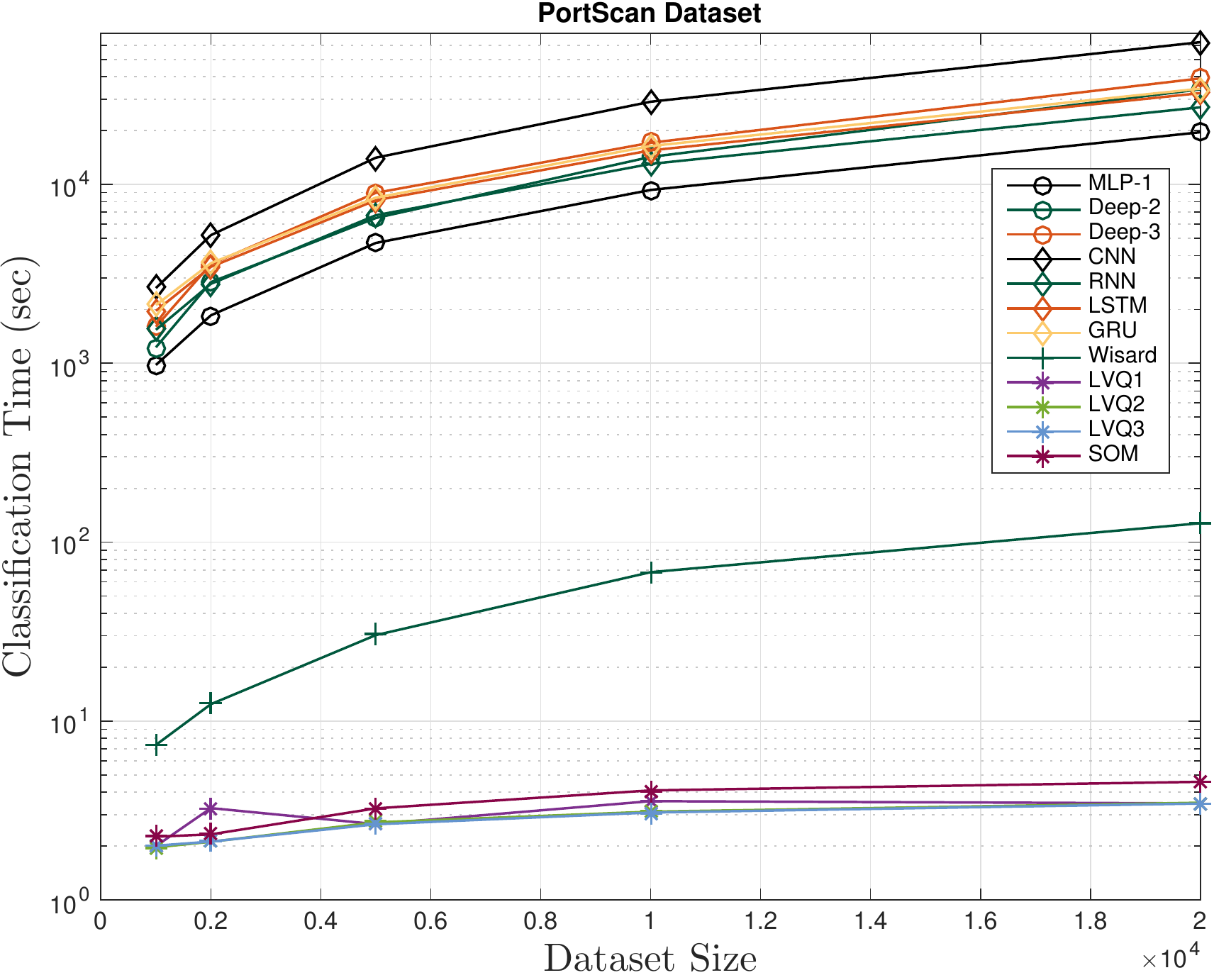}}   \hspace{2.5mm}
	\end{tabular}
	\caption{Single-class (DDoS): (a) Average Accuracy vs. dataset size;  (b) Classification Time vs. dataset size; Single-class (Portscan): (c) Average Accuracy vs. dataset size;  (d) Classification Time vs. dataset size.}
	\label{fig:aveacc_time}
\end{figure*}

\subsection{Single-Class Analysis}

We start by evaluating the performance of the different neural-based techniques by considering four single-class datasets ({\em DDoS}, {\em Portscan}, {\em WebAttack}, and {\em TOR}), reporting accuracy and F-measure in Figures \ref{fig:perf_single}(a), \ref{fig:perf_single}(b),  \ref{fig:perf_single}(c), and  \ref{fig:perf_single}(d), respectively. The choice of four completely different datasets (representative of four substantially different network attacks) is useful to verify the effectiveness of the tested algorithms and their relationships with the data.

Among classic ANN-based algorithms we distinguish between deep versions (Deep-$2$, Deep-$3$) and non-deep ones (MLP-$1$) whose detailed structure is reported in Table \ref{tab:hyperparams}. MLP-$1$ refers to a standard MLP with $3$ layers: $1$ input layer, $1$ hidden layer, and $1$ output layer. On the other hand, deep versions are indicated by Deep-$2$ ($1$ input layer, $2$ hidden layer, $1$ output layer) and Deep-$3$ ($1$ input layer, $3$ hidden layer, $1$ output layer). In order to compare algorithms with similar performances, MLP-$1$, Deep-$2$ and Deep-$3$ have been set with the same number of weights ($2k$). This is about one order of magnitude smaller than the dataset size, according to what is recommended in the literature \cite{lisboa,prasad08}. We also note that the number of weights for MLP-$1$, Deep-$2$ and Deep-$3$ directly results from the different number of neurons chosen per each hidden layer, namely: $26$ neurons for MLP-1, $23$ and $10$ neurons for the two hidden layers of Deep-2, and $20$, $16$, and $11$ neurons for the $3$ hidden layers of Deep-3 (see also Table \ref{tab:hyperparams}). 
As concerns the evolved deep models (CNN, RNN, LSTM, GRU), details about their structures are available in Table \ref{tab:hyperparams}. Also for these approaches, the choice of a particular structure (e.g. the number of filters in CNN, the number of recurrent units in RNN, and so forth), is aimed at obtaining a number of weights comparable with MLP-$1$, Deep-$2$, and Deep-$3$ so to have a fair comparison. 
	
In our analysis, the MLP-based versions (MLP-$1$, Deep-$2$ and Deep-$3$) and the evolved deep techniques (CNN, RNN, LSTM, GRU) exhibit good values of average accuracy (all around $0.99$) for all the four examined datasets (panels of Fig. \ref{fig:perf_single}).
Similarly, for all the aforementioned algorithms and for each dataset, F-measure exhibits high values, indicating a very negligible fraction of false positives and false negatives. 
Similar high performance is exhibited by Wisard for all the datasets with a slight exception of Portscan Dataset.
%Among the remaining algorithms, WiSARD seems to offer the best overall performance with the DDoS dataset (average accuracy amounting to $=0.9953$). With the Portscan dataset, WiSARD exhibits lower performance values (average accuracy amounting to $=0.8901$); but it exhibits smaller f-measure oscillations if compared with LVQs and SOM. Again, for the WebAttack and the TOR datasets, Wisard has an average accuracy amounting to $0.9894$ and $0.997$, respectively. 
By contrast, LVQs and SOM methodologies have lower performance and seem to suffer from the presence of more false positives and false negatives. This is visible for the Portscan dataset (Fig. \ref{fig:perf_single}(b)) that exhibits an oscillating F-measure, which can be ascribed to the different codebook vectors obtained when considering different input datasets.
In particular, competitive learning methods seem suffer from the particular type of attack dataset. In fact, the Euclidean distance (which is used as the main metric within the discussed competitive learning approaches) may not perfectly capture the properties of various datasets, being they structurally different. This is due to the fact that Euclidean distance is a not scale invariant measure, thus, it has difficulty to deal with vector components having different dynamic range. This is also the reason why the $3$ variants of LVQ can behave slightly differently when considering diverse datasets. 
	
%In summary, MLP-based algorithms outperform other techniques in terms of accuracy/F-measure figures. Wisard is able to hold a candle (with the slight exception of Portscan Dataset), whereas LVQs and SOM exhibit less (but not dramatic) performance.   

Another useful analysis is aimed at evaluating the impact of dataset size on performance. For the sake of compactness, we choose DDoS and Portscan as benchmark datasets, being DDoS and Portscan representative of two completely different and well structured network attacks.
Precisely, we evaluate the average accuracy (between DDoS/Portscan and benign classes) against an interval of dataset size varying between $10^3$ and $2\cdot10^4$ instances. The results shown in Figures \ref{fig:aveacc_time}(a) and \ref{fig:aveacc_time}(c) provide a clear indication that: \textit{i)} both ANNs and WiSARD exhibit very stable performance as dataset size varies; \textit{ii)} LVQ shows some minor fluctuations, with an average accuracy that is just slightly under $0.9$ in few cases. In this case, fluctuations can be ascribed to the codebook size that would require heuristical adjustments as the dataset size varies.

For the sake of fairness, the performance comparison must be complemented by a time-complexity analysis, as shown in Figs. \ref{fig:aveacc_time}(b-d). Therein, classification time (in seconds) is plotted against the number of instances (from $10^3$ to $2\cdot10^4$). For both experiments (DDoS dataset in \ref{fig:aveacc_time}(b) and Portscan dataset in \ref{fig:aveacc_time} (d)) we can reasonably recognize three slots in the Y axis: the first one, ranging from $10^3$ to $15\cdot10^4$ seconds, where MLP-$1$, Deep-$2,3$ and evolved deep architectures (CNN, RNN, LSTM, GRU) operate (slowest algorithms); the second one, ranging from about $10$ to $10^2$ seconds which includes WiSARD (medium-fast algorithm); the third one, ranging from about $1$ to $5$ seconds, including LVQ$1$, LVQ$2$, LVQ$3$, and SOM (faster algorithms). 

We are not surprised that  deep networks are up to $2$ orders of magnitude slower w.r.t. other techniques, due to the fully connected neurons between each layer. As regards the evolved deep architectures, time complexity is typically associated to the adoption of more sophisticated structures such as the convolutional layers (CNNs), or mechanisms to retain the internal memory states (RNN, LSTM, GRU). 
	
\begin{figure*}[t!] 
	\centering
	\begin{tabular}{cc}
		\subfloat[]{\includegraphics[scale=0.51]{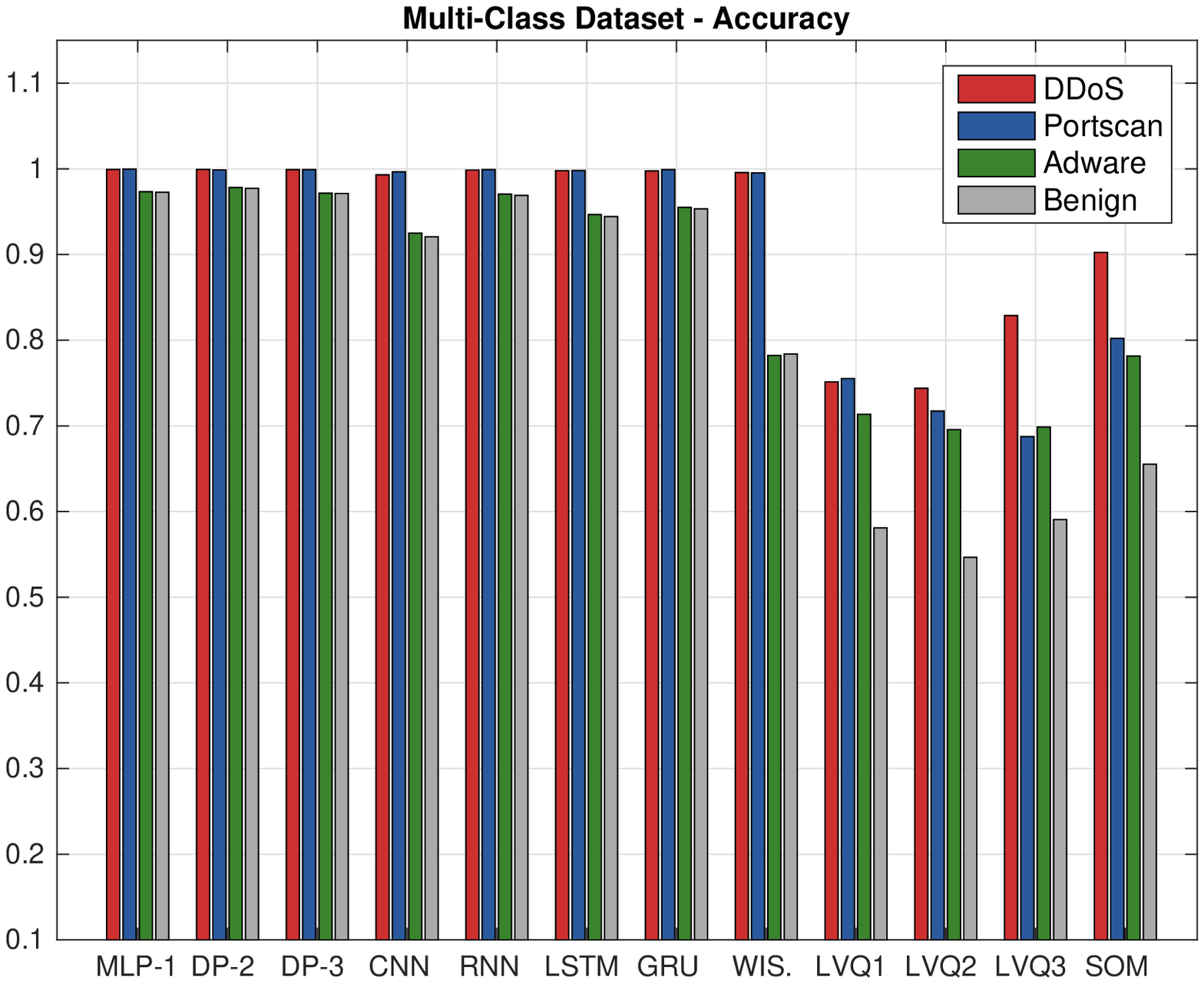}}  \hspace{2.5mm}
		\subfloat[]{\includegraphics[scale=0.51]{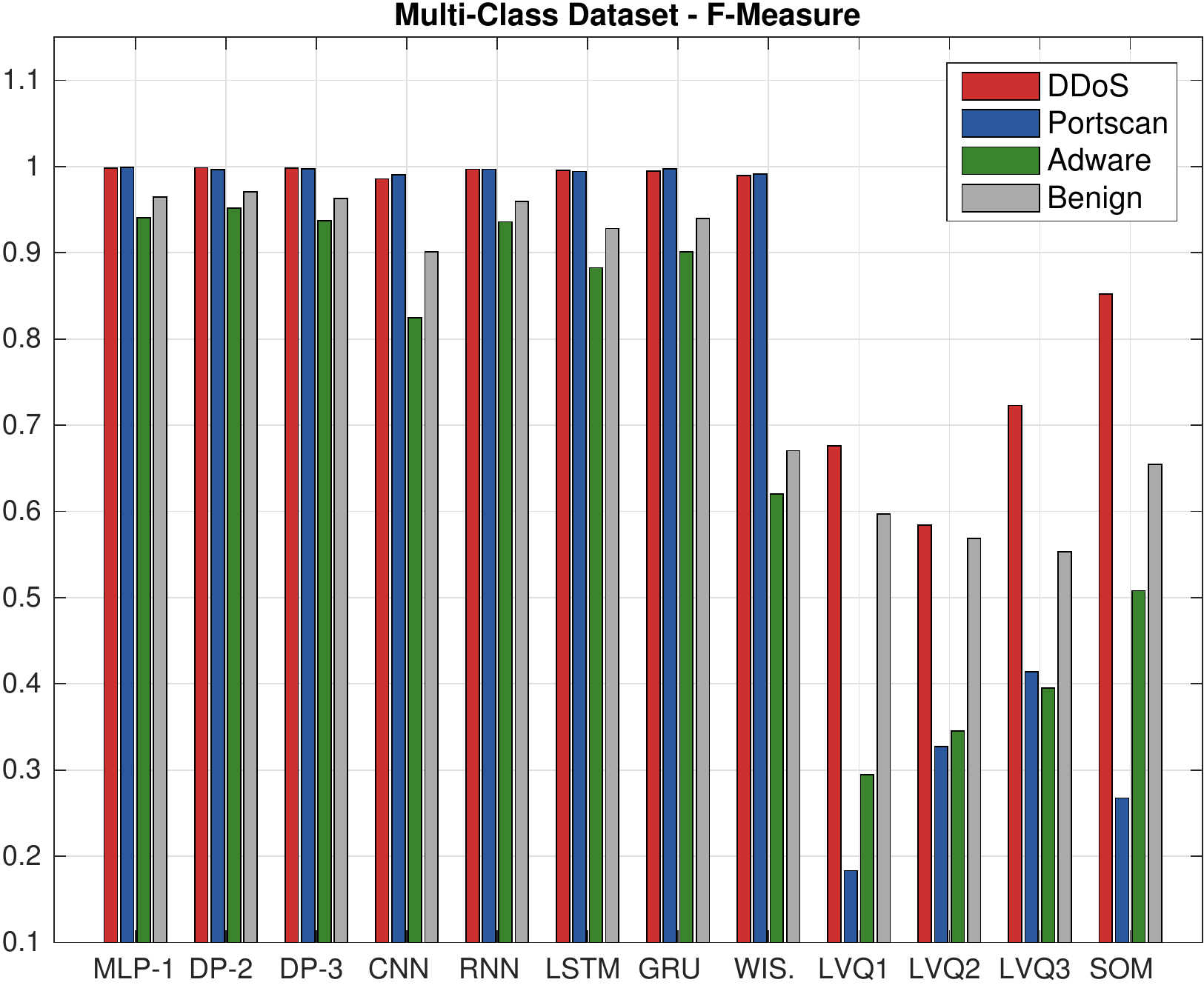}}  
	\end{tabular}
	\caption{Performance analysis for a $4$-classes dataset: (a) Accuracy, (b) F-Measure.}
	\label{fig:perf_multi}
\end{figure*}
\begin{figure*}[t!] 
	\centering
	\begin{tabular}{cc}
		\subfloat[]{\includegraphics[scale=0.5]{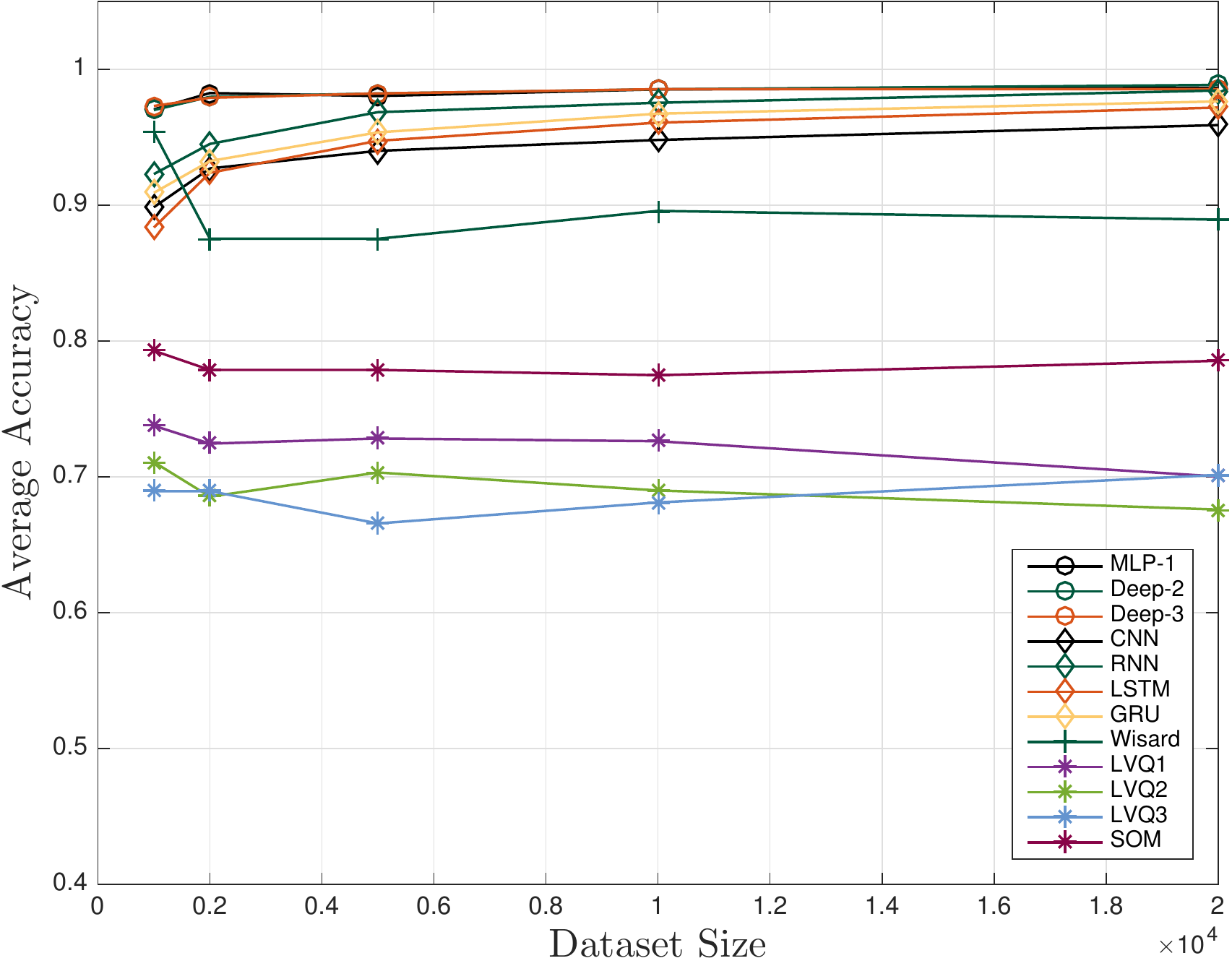}}  \hspace{2.5mm}
		\subfloat[]{\includegraphics[scale=0.5]{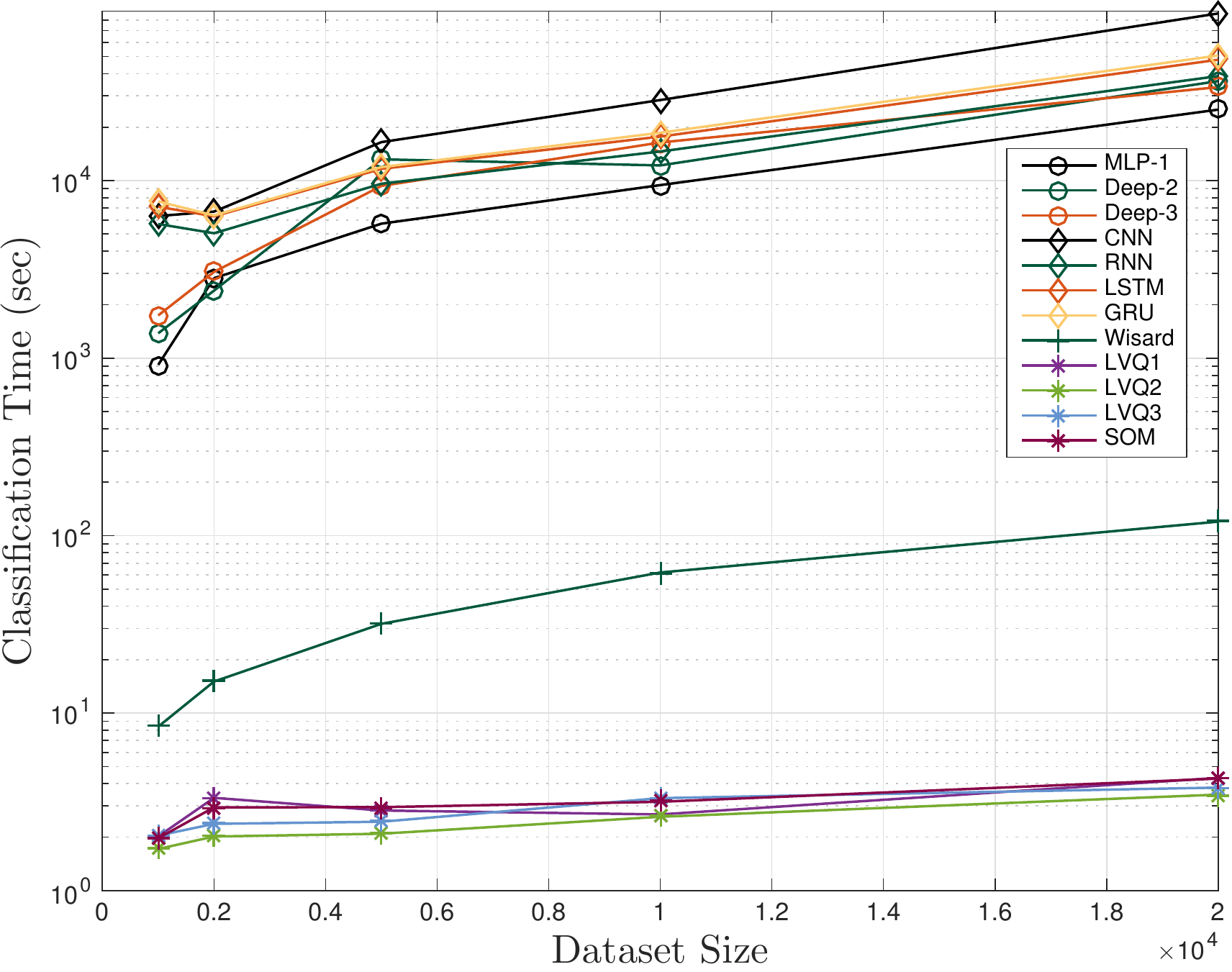}}  
	\end{tabular}
	\caption{Multi-class dataset: (a) Average Accuracy vs. dataset size; \\ ~~~ (b) Classification Time vs. dataset size.}
	\label{fig:multi_aveacc_time}
\end{figure*}
On the other hand, approaches based on competitive learning (LVQs, SOM) are faster since only the winner neurons are enabled to update their weights, although this comes at the cost of a lower accuracy. 
Surprisingly, the best trade-off between accuracy and time complexity is offered by WiSARD. Its RAM-based structure, in fact, allows a fast learning stage since it deals with binary information regardless on data dimension.  
In terms of time complexity, the feed-forward incremental learning scheme implemented in WiSARD is more advantageous than classic backpropagation, requiring many iterations to converge. 
\begin{figure*}[t!] 
	\centering
	\begin{tabular}{cc}
		\subfloat[]{\includegraphics[scale=0.38,angle=90]{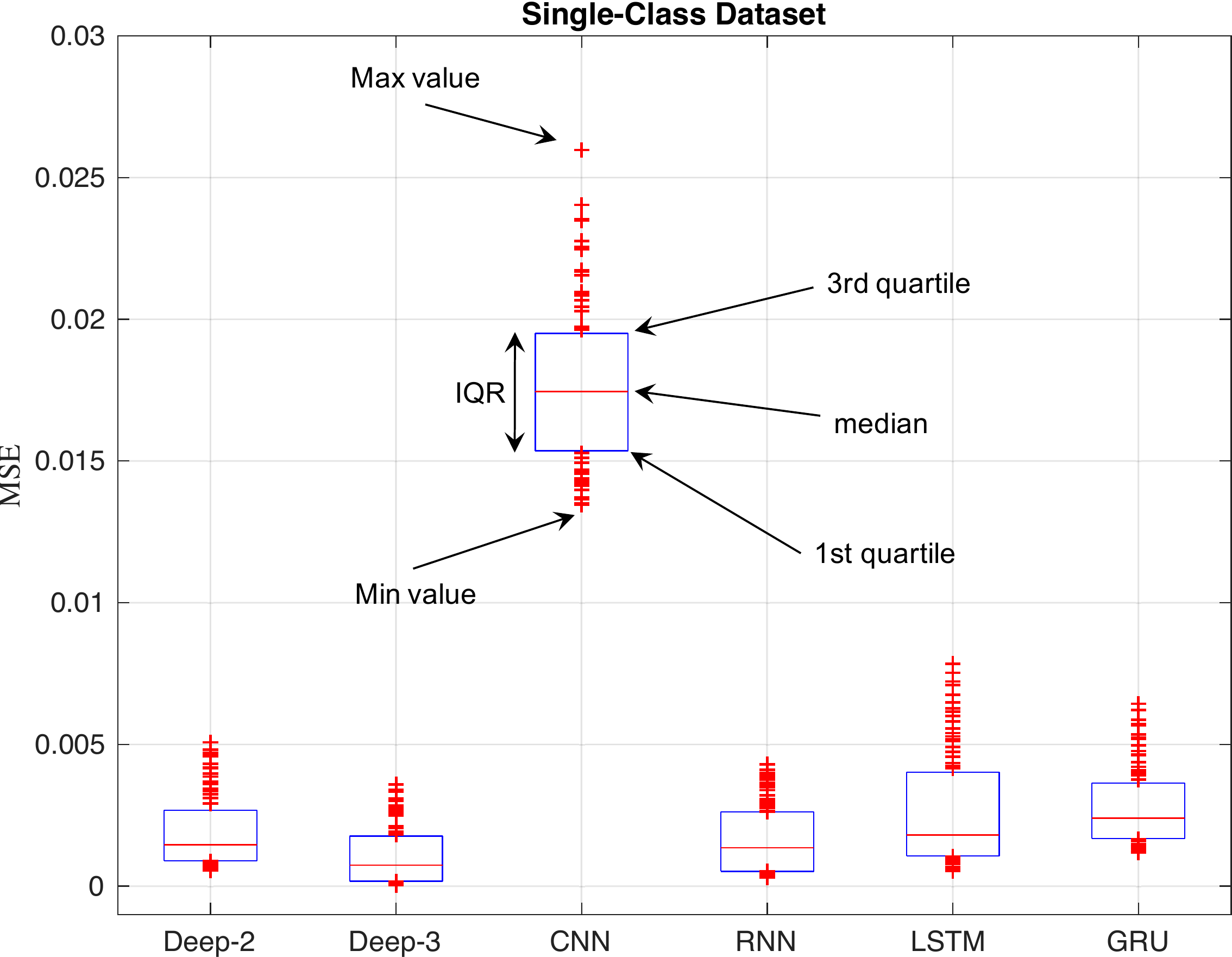}}  \hspace{3mm}
		\subfloat[]{\includegraphics[scale=0.5]{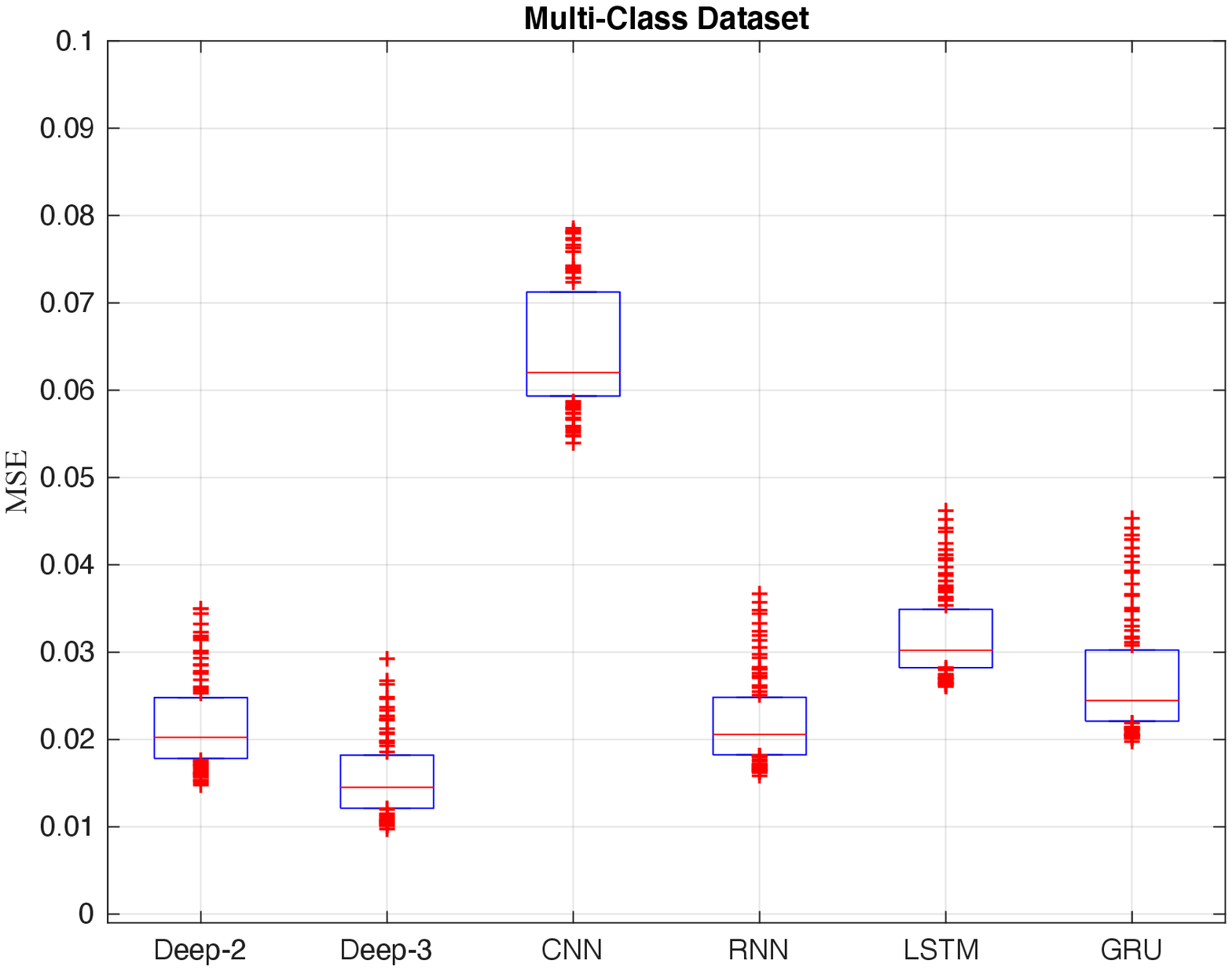}}  
	\end{tabular}
	\caption{MSE analysis when introducing more sophisticated deep models (CNN, RNN, LSTM, GRU). Single-class (DDoS benchmark dataset) (a), Multi-class (DDoS-Portscan-Adware-Benign dataset) (b).}
	\label{fig:mse_a2}
\end{figure*}

% Table generated by Excel2LaTeX from sheet 'Foglio1'

%\vspace{-20pt}
\subsection{Multi-Class Analysis}

Multi-class analysis is useful to analyze how the various algorithms react when dealing with different classes of traffic. With this aim, we built a $4$-class dataset by mixing (in a balanced way) three malicious classes (DDoS, Portscan, and Adware) with a Benign class. The rationale behind this choice is to take into account a mix of peculiar threats.
As regards the performance analysis, for the sake of readiness, we show multi-class accuracy and F-measure separately in Figs. \ref{fig:perf_multi}(a) and \ref{fig:perf_multi}(b), respectively. As a general trend, we observe a satisfactory classification performance by MLP and Deep algorithms. WiSARD exhibits good performance in classifying DDoS and Portscan ($0.9957$ and $0.9952$ accuracy, respectively), but is poor in  classifying Adware and Benign classes ($0.784$ and $0.782$  accuracy, respectively). The reason is to be found behind the structural difference between those different types of attacks. DDoS and Portscan are both highly ``structured" attacks. The former can be characterized in terms of high rate of messages that a bot sends to a victim; Portscan hides continuous ping-based requests towards a target in order to unveil possible open ports. By contrast, Adwares can be easily confused within a benign flow, since they just convey annoying banners, as often occurs also within legitimate web portals. 
The situation changes drastically when we look at the remaining algorithms (LVQs and SOM), where false positives have high weight especially for Portscan, Adware, and Benign classes, as shown in Fig. \ref{fig:perf_multi}(b). Also, the F-measure is often below $0.4$ in case of Adware and Benign classes. With respect to the single-class case, here, the effects produced by the codebook class assignment and peculiarity of Euclidean measure are amplified.

Finally, among competitive-based approaches, LVQ-$3$ exhibits the best performance, due to the introduction of the empirical parameter $\epsilon$ (see Sect. \ref{sec:lvq3}), which helps to better regulating the distance between the data and the pertinent codebook vectors.
This notwithstanding, all the considered techniques exhibit fairly stable accuracy averages, across the whole range of dataset sizes (from $10^3$ to $2\cdot10^4$ instances), as revealed by Fig. \ref{fig:multi_aveacc_time}(a). As expected, the average accuracy in all cases pertaining the multi-class experiment is slightly lower than their single-class counterpart, due to the fact that classifying more than two classes is a more challenging task.  

Let us now discover if the multi-class analysis has an impact on the time complexity of the considered algorithms. Figure  \ref{fig:multi_aveacc_time}(b) shows how the order of magnitude in classification times per algorithm is similar w.r.t. the case of single-class analysis. As to be expected, the slight increase in classification time going from single to multi-class cases  can be ascribed to the higher number of different traffic flows to be classified. For example, along the varying dataset size $\left[ 10^3, 2\cdot10^3, 5\cdot10^3, 10^4, 2\cdot10^4 \right]$, WiSARD exhibits the following (approximated) classification times: $\left[ 8.047, 12.24, 30.731, 53.474, 104.783  \right]$ seconds in the single class case, and $\left[ 8.389, 15.058, 31.831, 61.958, 119.848 \right]$ seconds in the multi-class case. Once again, WiSARD appears to provide the optimal trade-off between performance needs and time complexity issues.

\subsection{General Considerations}
A number of interesting considerations may be derived from our comparative analysis. First, not all neural-based techniques are equally applicable to network intrusion detection, particularly when performance is the key issue. 

ANN approaches (including modern deep-based methods) tend to offer noteworthy accuracy, both in single-class and multi-class datasets. Unfortunately, though, accuracy comes at the expenses of time complexity, which will demand for deep-based techniques to rely on specialized GPU-based hardware.

Moreover, as an auxiliary investigation pertaining the deep approach, we evaluate the impact of the updating weight mechanism on the overall performance, expressed through the Mean Square Error (MSE). The behavior is shown in Figs. \ref{fig:mse_a2}(a) and \ref{fig:mse_a2}(b) for the single and multi-class datasets, respectively, evaluated over $100$ epochs. Such analysis reveals that most techniques exhibit limited fluctuations around the median value of MSE and a quite low MSE value. The only exception is given by CNN where the higher MSE is reasonably due to the complexity of the convolution operation. Similar considerations raised in a bio-informatic study \cite{cnn_fluct} where only MLP and CNN techniques have been compared. Detailed values are reported in Table \ref{tab:mse_results}, in terms of median and inter-quartile range (difference between third and first quartile) values. On average, the most ``stable'' techniques are the classic Deep-$2$ and Deep-$3$ architectures. The addition of a bit more sophisticated structure (e.g. convolutional layer for CNN, LSTM/GRU units, and so forth) can translate into MSE fluctuations due to the presence of additional hyperparameters to tune. Obviously, such fluctuations directly reflect the internal structure of each deep technique and can be more (e.g. CNN) or less (e.g. RNN) pronounced.

\begin{table}[ht]
		\centering
		\caption{ Median and IQR values (MSE analysis) for deep-based techniques}
		\begin{tabular}{c  c c  c c }
			& \multicolumn{2}{c}{\textbf{Median}} & \multicolumn{2}{c}{\textbf{IQR}} \\
			\cmidrule{1-5}    \textbf{Technique} & \multicolumn{1}{c}{Single-Class} & \multicolumn{1}{c}{Multi-Class} & \multicolumn{1}{c}{Single-Class} & \multicolumn{1}{c}{Multi-Class} \\
			\midrule
			Deep-2 & 1.5$\cdot$ 10$^{-3}$ & 2.02 $\cdot$ 10$^{-2}$ & 1.8$\cdot$ 10$^{-3}$ & 7 $\cdot$ 10$^{-3}$ \\
			Deep-3 & 7.4$\cdot$ 10$^{-4}$ & 1.45 $\cdot$ 10$^{-2}$ & 1.6$\cdot$ 10$^{-3}$ & 6.1$\cdot$ 10$^{-3}$ \\
			CNN   & 1.74 $\cdot$ 10$^{-2}$ & 6.2 $\cdot$ 10$^{-2}$ & 4.1$\cdot$ 10$^{-3}$ & 1.19$\cdot$ 10$^{-2}$ \\
			RNN   & 1.4 $\cdot$ 10$^{-3}$ & 2.06 $\cdot$ 10$^{-2}$ & 2.1$\cdot$ 10$^{-3}$ & 6.6$\cdot$ 10$^{-3}$ \\
			LSTM  & 1.8 $\cdot$ 10$^{-3}$ & 3.02 $\cdot$ 10$^{-2}$ & 2.9$\cdot$ 10$^{-3}$ & 6.7 $\cdot$ 10$^{-3}$ \\
			GRU   & 2.4 $\cdot$ 10$^{-3}$ & 2.45 $\cdot$ 10$^{-2}$ & 2$\cdot$ 10$^{-3}$ & 8.1 $\cdot$ 10$^{-3}$ \\ \hline
		\end{tabular}%
		\label{tab:mse_results}%
\end{table}%

Remarkably, deep-learning techniques could find interesting application in the network intrusion management field when the training set exhibits slow time dynamics. In this case, the training operation (being time/resource consuming) could be performed only once (or performed rarely across the time). 

As regards competitive learning techniques (LVQ in the $3$ variants and SOM), pros and cons are (almost) reversed w.r.t. ANN algorithms. In the case of single-class analysis, a dependency on the type of dataset  is highlighted (better performance in DDoS dataset than in Portscan). This is reasonably due to the fact that different inputs produce different mappings with codebook vectors. As to be expected, this effect is further amplified in multi-class datasets, where false positives and false negatives lead to unstable f-measure figures (Fig. \ref{fig:perf_multi} (b)). 

However, the competitive learning algorithms exhibit very appealing time complexity figures (up to $3$ orders of magnitude lower than ANNs), thanks to their approach to taking into account only the ``winner" neurons during the whole classification process. Accordingly, cooperative techniques could find useful application in highly time-variant intrusion detection settings, where it is crucial to quickly adapt to the dataset variations. Possibly, they can operate an early (even rough) detection to be refined later by means of other algorithms.

Finally, an unexpected finding is offered by WiSARD that, through its weightless mechanism, provides the best trade-off between performance (accuracy/F-measure) and time complexity, in both single and multi-class cases. This outstanding behavior is mainly due to two aspects: first, the binarization scheme (jointly with the RAM-like neurons structure) adopted by WiSARD allows to handle multivariable data in a fast way. Then, the training stage
follows an incremental approach since each sample reinforces the past knowledge in updating the network state; this implies a fast convergence and makes WiSARD particularly suitable in domains where  online learning is crucial.
What is even more interesting is that WiSARD has hardly ever been considered in the field of traffic flow classification, while it appears to be a very interesting candidate to be exploited in intrusion detection. 
This WNN-based approach should certainly be considered as a promising alternative to existing methods, particularly for its potential to meet the near-real-time constraints involved in NIDS classification problems.

\section{Conclusion}
\label{sec:conclusion}

This work explores the applicability of prominent neural-based techniques in network intrusion detection. We carry out an experiment-based comparison to quantify performance and trade-off achievable with each solution. Our aim to perform a fair comparison has directed our investigation to focus on artificial neural networks (ANN). This was to avoid the typical issues arising when comparing ML methods relying on different rationale (e.g. ANN vs. SVM vs. Decision Trees).

In the related work section, we have provided pointers to numerous other surveys, illustrating similarities (in aims) and differences (in methodologies). The key peculiarity of our paper is its experimental-based approach to reviewing alternative ANN options. We based our evaluation on modern datasets (CIC-IDS-2017/2018), while most of the earlier experimental results are based on the outdated KDD$99$ dataset (reflecting network attack issues that have largely been solved). Our review provides useful performance (in terms of accuracy and F-measure) and time complexity data, providing the basis for a trade-off analysis across different ANNs such as deep networks or competitive learning-based networks. To add value to our study, we wanted to consider also methods that have not typically been employed in intrusion detection, particularly the weightless neural networks (WNNs).

The outcomes reveal a number of interesting findings. \textit{i)} ANN-based approaches (including deep networks) are characterized by outstanding performance in almost all cases (as to be expected). Yet, they suffer the drawback of being slow due to the underlying backpropagation algorithm, which is typically slow to converge.  \textit{ii)} Neural-based techniques relying on a competitive learning approach (LVQ$1$, LVQ$2$, LVQ$3$, SOM) overturn this perspective, thanks to a much reduced time complexity (due to the winner-neuron mechanism). However, this comes with  a lower performance when different datasets are considered (due to the different mapping mechanism between input and codebook vectors). \textit{iii)} The WiSARD algorithm (representative of WNNs) exhibits a surprisingly best trade-off in terms of performance and time complexity, making it an appealing candidate to operate in conjunction with intrusion detection systems.   

The proposed analysis could be extended in several ways. 
The first (and perhaps natural) direction is to better investigate the potential of weightless neural networks in the context of network intrusion detection. Beyond WiSARD, in fact, other WNN algorithms such as Probabilistic Logic Node (PLN), Goal Seeking Neuron (GSN), or General Neural Unit (GNU) could reveal noteworthy properties when applied in this field. 
Then, we would suggest to explore how the addition of a feature selection (FS) pre-processing stage would reduce time complexity. This could be particularly advantageous for deep techniques (the most critical in terms of time constraints), even if the adoption of a non well designed FS strategy could result in a further computational overhead. 
Finally, having identified pros and cons of neural-based techniques in the field of intrusion detection, an engine which allows to automatically select (or combine) the NN-based strategies best fitting the underlying network environment (e.g. hugely vs. scarcely time-variant traffic profiles) can be designed. This latter point could have intriguing implications onto prospective $6G$ scenarios which, according to the network experts, will be characterized by automatic service provisioning, intelligent resource management, and smart network adjustments.

\bibliographystyle{unsrt}
\bibliography{ml}

\begin{thebibliography}{10}

\bibitem{granville1}
I.~Possebon, A.~Santos~da Silva, L.~Zambenedetti~Granville, A~Schaeffer-Filho,
  and A.K. Marnerides.
\newblock Improved network traffic classification using ensemble learning.
\newblock In {\em 2019 {IEEE} Symposium on Computers and Communications}, pages
  1--6, 2019.

\bibitem{granville2}
F.~Grando, L.~Zambenedetti~Granville, and L.C. Lamb.
\newblock Machine learning in network centrality measures: Tutorial and
  outlook.
\newblock {\em ACM Comput. Surv.}, 51(5):102:1--102:32, 2018.

\bibitem{cisco-nids}
Cisco {E}ncrypted {T}raffic {A}nalysis.
\newblock
  \url{https://www.cisco.com/c/dam/en/us/solutions/collateral/enterprise-networks/enterprise-network-security/nb-09-encrytd-traf-anlytcs-wp-cte-en.pdf}.
\newblock Accessed: 2020-02-25.

\bibitem{gaspary1}
L.~P. {Gaspary}, E.~{Meneghetti}, and L.~R. {Tarouco}.
\newblock An {SNMP} agent for stateful intrusion inspection.
\newblock In {\em IFIP/IEEE Eighth International Symposium on Integrated
  Network Management, 2003.}, pages 3--16, 2003.

\bibitem{gaspary2}
L.~P. {Gaspary}, R.~N. {Sanchez}, D.~W. {Antunes}, and E.~{Meneghetti}.
\newblock A {SNMP} -based platform for distributed stateful intrusion detection
  in enterprise networks.
\newblock {\em IEEE Journal on Selected Areas in Communications},
  23(10):1973--1982, 2005.

\bibitem{cerroni1}
W.~Cerroni, G.~Moro, R.~Pasolini, and M.~Ramilli.
\newblock Decentralized detection of network attacks through p2p data
  clustering of {SNMP} data.
\newblock {\em Computers \& Security}, 52:1 -- 16, 2015.

\bibitem{cerroni2}
W.~Cerroni, G.~Moro, R.~Pasolini, and M.~Ramilli.
\newblock Network attack detection based on peer-to-peer clustering of {SNMP}
  data.
\newblock In {\em Lecture Notes of the Institute for Computer Sciences},
  volume~22, 2009.

\bibitem{cerroni3}
W.~Cerroni, G.~Moro, T.~Pirini, and M.~Ramilli.
\newblock Peer-to-peer data mining classifiers for decentralized detection of
  network attacks.
\newblock In {\em Proceedings of the Twenty-Fourth Australasian Database
  Conference - Volume 137}, pages 101--107, 2013.

\bibitem{kdd-dataset}
Kdd cup 1999 data.
\newblock \url{http://kdd.ics.uci.edu/databases/kddcup99/kddcup99.html}.
\newblock Accessed: 2020-02-25.

\bibitem{nsl-kdd}
M.~{Tavallaee}, E.~{Bagheri}, W.~{Lu}, and A.~A. {Ghorbani}.
\newblock A detailed analysis of the {KDD CUP} 99 data set.
\newblock In {\em 2009 IEEE Symposium on Computational Intelligence for
  Security and Defense Applications}, pages 1--6, 2009.

\bibitem{icissp-dataset3}
{Sharafaldin I.}, {Habibi Lashkari A.}, and {Ghorbani A.A.}
\newblock Toward generating a new intrusion detection dataset and intrusion
  traffic characterization.
\newblock In {\em 4th International Conference on Information Systems Security
  and Privacy}, 2018.

\bibitem{classif_dimauro}
M.~Di Mauro and C.~Di Sarno.
\newblock Improving siem capabilities through an enhanced probe for encrypted
  {S}kype traffic detection.
\newblock {\em Journal of Information Security and Applications}, 38:85--95,
  2018.

\bibitem{classif_liotta}
F.~Cauteruccio, G.~Fortino, A.~Guerrieri, A.~Liotta, D.C. Mocanu, C.~Perra,
  G.~Terracina, and M.~Torres~Vega.
\newblock Short-long term anomaly detection in wireless sensor networks based
  on machine learning and multi-parameterized edit distance.
\newblock {\em Information Fusion}, 52:13 -- 30, 2019.

\bibitem{tnsm-classif-iot}
I.~{Hafeez}, M.~{Antikainen}, A.~Y. {Ding}, and S.~{Tarkoma}.
\newblock {I}o{T}-keeper: Detecting malicious {I}o{T} network activity using
  online traffic analysis at the edge.
\newblock {\em IEEE Transactions on Network and Service Management}, pages
  1--1, 2020.

\bibitem{classif_pascale}
M.~Casillo, S.~Coppola, M.~De~Santo, F.~Pascale, and E~Santonicola.
\newblock Embedded intrusion detection system for detecting attacks over
  can-bus.
\newblock In {\em 4th International Conference on System Reliability and
  Safety}, pages 136--141, 2019.

\bibitem{precic_paper2}
R.~U. {Khan}, X.~{Zhang}, M.~{Alazab}, and R.~{Kumar}.
\newblock An improved convolutional neural network model for intrusion
  detection in networks.
\newblock In {\em 2019 Cybersecurity and Cyberforensics Conference}, pages
  74--77, 2019.

\bibitem{precic_paper3}
S.~T.~F. {Al-Janabi} and H.~A. {Saeed}.
\newblock A neural network based anomaly intrusion detection system.
\newblock In {\em Developments in E-systems Engineering}, pages 221--226, 2011.

\bibitem{ids_ann_19}
K.~A. {Taher}, B.~{Mohammed Yasin Jisan}, and M.~M. {Rahman}.
\newblock Network intrusion detection using supervised machine learning
  technique with feature selection.
\newblock In {\em 2019 International Conference on Robotics,Electrical and
  Signal Processing Techniques (ICREST)}, pages 643--646, 2019.

\bibitem{precic_paper1}
S.~{Kumar} and A.~{Yadav}.
\newblock Increasing performance of intrusion detection system using neural
  network.
\newblock In {\em IEEE International Conference on Advanced Communications,
  Control and Computing Technologies}, pages 546--550, 2014.

\bibitem{precic_paper4}
D.~{Papamartzivanos}, F.~{Gomez Marmol}, and G.~{Kambourakis}.
\newblock Introducing deep learning self-adaptive misuse network intrusion
  detection systems.
\newblock {\em IEEE Access}, 7:13546--13560, 2019.

\bibitem{precic_paper5}
Z.~T. {Fernando}, I.~S. {Thaseen}, and C.~A. {Kumar}.
\newblock Network attacks identification using consistency based feature
  selection and self organizing maps.
\newblock In {\em First International Conference on Networks Soft Computing},
  pages 162--166, 2014.

\bibitem{precic_paper6}
S.~{McElwee} and J.~{Cannady}.
\newblock Improving the performance of self-organizing maps for intrusion
  detection.
\newblock In {\em SoutheastCon 2016}, pages 1--6, 2016.

\bibitem{precic_paper7}
C.~{Li-ying}, Z.~{Xiao-xian}, L.~{He}, and C.~{Gui-fen}.
\newblock A network intrusion detection method based on combined model.
\newblock In {\em International Conference on Mechatronic Science, Electric
  Engineering and Computer}, pages 254--257, 2011.

\bibitem{precic_paper8}
Z.N.Al-Sultani R.S.Naoum.
\newblock Learning vector quantization (lvq) and k-nearest neighbor for
  intrusion classification.
\newblock {\em World of Computer Science and Information Technology Journal},
  2(3):105--109, 2012.

\bibitem{ids_dnn_1}
J.~{Woo}, J.~{Song}, and Y.~{Choi}.
\newblock Performance enhancement of deep neural network using feature
  selection and preprocessing for intrusion detection.
\newblock In {\em 2019 International Conference on Artificial Intelligence in
  Information and Communication (ICAIIC)}, pages 415--417, 2019.

\bibitem{ids_dnn_2}
Y.~{Jia}, M.~{Wang}, and Y.~{Wang}.
\newblock Network intrusion detection algorithm based on deep neural network.
\newblock {\em IET Information Security}, 13(1):48--53, 2019.

\bibitem{ids_dnn3}
A.~{Nagisetty} and G.~P. {Gupta}.
\newblock Framework for detection of malicious activities in {I}o{T} networks
  using {K}eras deep learning library.
\newblock In {\em 2019 3rd International Conference on Computing Methodologies
  and Communication (ICCMC)}, pages 633--637, 2019.

\bibitem{ids_dnn4}
F.~A. {Khan}, A.~{Gumaei}, A.~{Derhab}, and A.~{Hussain}.
\newblock A novel two-stage deep learning model for efficient network intrusion
  detection.
\newblock {\em IEEE Access}, 7:30373--30385, 2019.

\bibitem{kdd_svm1}
Z.~{Zhang} and P.~{Pan}.
\newblock A hybrid intrusion detection method based on improved fuzzy c-means
  and support vector machine.
\newblock In {\em 2019 International Conference on Communications, Information
  System and Computer Engineering (CISCE)}, pages 210--214, 2019.

\bibitem{kdd_svm2}
W.~{Wang}, X.~{Du}, and N.~{Wang}.
\newblock Building a cloud ids using an efficient feature selection method and
  svm.
\newblock {\em IEEE Access}, 7:1345--1354, 2019.

\bibitem{kdd_svm3}
X.~{Tang}, S.~X.~. {Tan}, and H.~{Chen}.
\newblock {SVM} based intrusion detection using nonlinear scaling scheme.
\newblock In {\em 2018 14th IEEE International Conference on Solid-State and
  Integrated Circuit Technology (ICSICT)}, pages 1--4, 2018.

\bibitem{kdd_svm4}
S.~{Sun}, Z.~{Ye}, L.~{Yan}, J.~{Su}, and R.~{Wang}.
\newblock Wrapper feature selection based on lightning attachment procedure
  optimization and support vector machine for intrusion detection.
\newblock In {\em (IDAACS-SWS)}, pages 41--46, 2018.

\bibitem{kdd_svm5}
S.~{Teng}, N.~{Wu}, H.~{Zhu}, L.~{Teng}, and W.~{Zhang}.
\newblock Svm-dt-based adaptive and collaborative intrusion detection.
\newblock {\em IEEE/CAA Journal of Automatica Sinica}, 5(1):108--118, 2018.

\bibitem{kdd_pca1}
A.~{Hadri}, K.~{Chougdali}, and R.~{Touahni}.
\newblock A network intrusion detection based on improved nonlinear fuzzy
  robust {PCA}.
\newblock In {\em 2018 IEEE 5th International Congress on Information Science
  and Technology (CiSt)}, pages 636--641, 2018.

\bibitem{kdd_pca2}
F.~{Meng}, Y.~{Fu}, F.~{Lou}, and Z.~{Chen}.
\newblock An effective network attack detection method based on kernel {PCA}
  and {LSTM-RNN}.
\newblock In {\em 2017 International Conference on Computer Systems,
  Electronics and Control (ICCSEC)}, pages 568--572, 2017.

\bibitem{kdd_pca3}
K.~{Ibrahimi} and M.~{Ouaddane}.
\newblock Management of intrusion detection systems based-{KDD99}: Analysis
  with {LDA} and {PCA}.
\newblock In {\em 2017 International Conference on Wireless Networks and Mobile
  Communications (WINCOM)}, pages 1--6, 2017.

\bibitem{kdd_pca4}
S.~M. {Almansob} and S.~S. {Lomte}.
\newblock Addressing challenges for intrusion detection system using naive
  {B}ayes and {PCA} algorithm.
\newblock In {\em 2017 2nd International Conference for Convergence in
  Technology (I2CT)}, pages 565--568, 2017.

\bibitem{kdd_pca5}
A.~A. {Aburomman} and M.~B.~I. {Reaz}.
\newblock Ensemble of binary {SVM} classifiers based on {PCA} and {LDA} feature
  extraction for intrusion detection.
\newblock In {\em {IEEE} (IMCEC)}, pages 636--640, 2016.

\bibitem{kdd_decision1}
R.~{Primartha} and B.~A. {Tama}.
\newblock Anomaly detection using random forest: A performance revisited.
\newblock In {\em 2017 International Conference on Data and Software
  Engineering (ICoDSE)}, pages 1--6, 2017.

\bibitem{kdd_decision2}
X.~{Gao}, C.~{Shan}, C.~{Hu}, Z.~{Niu}, and Z.~{Liu}.
\newblock An adaptive ensemble machine learning model for intrusion detection.
\newblock {\em IEEE Access}, 7:82512--82521, 2019.

\bibitem{kdd_decision3}
N.~{Kumar}, H.~{Akash}, R.~A. {Prataap}, G.~{Srinath}, and C.~{Mala}.
\newblock Intelligent intrusion detection system using decision tree classifier
  and bootstrap aggregation.
\newblock In {\em {ISED}}, pages 199--203, 2018.

\bibitem{kdd_decision4}
M.~{Bitaab} and S.~{Hashemi}.
\newblock Hybrid intrusion detection: Combining decision tree and gaussian
  mixture model.
\newblock In {\em 2017 14th International ISC (Iranian Society of Cryptology)
  Conference on Information Security and Cryptology (ISCISC)}, pages 8--12,
  2017.

\bibitem{kdd_decision5}
M.~A. {Jabbar} and S.~{Samreen}.
\newblock Intelligent network intrusion detection using alternating decision
  trees.
\newblock In {\em 2016 International Conference on Circuits, Controls,
  Communications and Computing (I4C)}, pages 1--6, 2016.

\bibitem{kdd_unsup1}
Z.~{Rustam} and A.~S. {Talita}.
\newblock Fuzzy kernel robust clustering for anomaly based intrusion detection.
\newblock In {\em 2018 Third International Conference on Informatics and
  Computing (ICIC)}, pages 1--4, 2018.

\bibitem{kdd_unsup2}
K.~{Alrawashdeh} and C.~{Purdy}.
\newblock Fast hardware assisted online learning using unsupervised deep
  learning structure for anomaly detection.
\newblock In {\em 2018 International Conference on Information and Computer
  Technologies (ICICT)}, pages 128--134, 2018.

\bibitem{kdd_unsup3}
M.~Z. {Alom} and T.~M. {Taha}.
\newblock Network intrusion detection for cyber security using unsupervised
  deep learning approaches.
\newblock In {\em 2017 IEEE National Aerospace and Electronics Conference
  (NAECON)}, pages 63--69, 2017.

\bibitem{kdd_unsup4}
W.~{Chen}, F.~{Kong}, F.~{Mei}, G.~{Yuan}, and B.~{Li}.
\newblock A novel unsupervised anomaly detection approach for intrusion
  detection system.
\newblock In {\em 3rd {I}nternational conference on big data security on
  cloud}, pages 69--73, 2017.

\bibitem{kdd_unsup5}
S.~{Seo}, S.~{Park}, and J.~{Kim}.
\newblock Improvement of network intrusion detection accuracy by using
  restricted {B}oltzmann machine.
\newblock In {\em 2016 8th International Conference on Computational
  Intelligence and Communication Networks (CICN)}, pages 413--417, 2016.

\bibitem{kdd_unsup6}
F.~Amiri, M.~Rezaei Yousefi, C.~Lucas, A.~Shakery, and N.~Yazdani.
\newblock Cluster ensemble with link-based approach for botnet detection.
\newblock {\em Journal of Network and Systems Management}, 26(3):616 -- 639,
  2018.

\bibitem{unsw2}
N.~{Moustafa} and J.~{Slay}.
\newblock {UNSW-NB15}: a comprehensive data set for network intrusion detection
  systems ({UNSW-NB15} network data set).
\newblock In {\em 2015 Military Communications and Information Systems
  Conference (MilCIS)}, pages 1--6, 2015.

\bibitem{bot-iot1}
N.{Koroniotis}, N.{Moustafa}, E.~{Sitnikova}, and B.{Turnbull}.
\newblock Towards the development of realistic botnet dataset in the {I}nternet
  of {T}hings for network forensic analytics: {B}o{T}-{I}o{T} dataset.
\newblock {\em Future Generation Computer Systems}, 100:779--796, 2019.

\bibitem{cic}
Canadian institute for cybersecurity.
\newblock \url{https://www.unb.ca/cic/}.
\newblock Accessed: 2020-02-25.

\bibitem{aws}
A realistic cyber defense dataset.
\newblock \url{https://registry.opendata.aws/cse-cic-ids2018/}.
\newblock Accessed: 2020-02-25.

\bibitem{cic_paper1}
V.~Kanimozhi and T.~P. Jacob.
\newblock Artificial intelligence based network intrusion detection with
  hyper-parameter optimization tuning on the realistic cyber dataset
  cse-cic-ids2018 using cloud computing.
\newblock In {\em 2019 International Conference on Communication and Signal
  Processing}, pages 33--36, 2019.

\bibitem{cic_paper2}
C.~{Ma}, X.~{Du}, and L.~{Cao}.
\newblock Analysis of multi-types of flow features based on hybrid neural
  network for improving network anomaly detection.
\newblock {\em IEEE Access}, 7:148363--148380, 2019.

\bibitem{cic_paper3}
S.~{Wankhede} and D.~{Kshirsagar}.
\newblock Do{S} attack detection using machine learning and neural network.
\newblock In {\em Fourth International Conference on Computing Communication
  Control and Automation}, pages 1--5, 2018.

\bibitem{cic_paper4}
J.~{Lee}, J.~{Kim}, I.~{Kim}, and K.~{Han}.
\newblock Cyber threat detection based on artificial neural networks using
  event profiles.
\newblock {\em IEEE Access}, 7:165607--165626, 2019.

\bibitem{survey_pure1}
T.~T.~T. {Nguyen} and G.~{Armitage}.
\newblock A survey of techniques for internet traffic classification using
  machine learning.
\newblock {\em IEEE Communications Surveys Tutorials}, 10(4):56--76, 2008.

\bibitem{survey_pure2}
R.~Boutaba, M.~A. Salahuddin, N.~Limam, S.~Ayoubi, N.~Shahriar,
  F.~Estrada-Solano, and O.~M. Caicedo.
\newblock A comprehensive survey on machine learning for networking: evolution,
  applications and research opportunities.
\newblock {\em Journal of Internet Services and Applications}, 9(1):16, 2018.

\bibitem{survey_pure3}
H.~Hindy, D.~Brosset, E.~Bayne, A.~Seeam, C.~Tachtatzis, R.~Atkinson, and
  X.~Bellekens.
\newblock A taxonomy and survey of intrusion detection system design
  techniques, network threats and datasets.
\newblock {\em CoRR}, abs/1806.03517, 2018.

\bibitem{survey_pure4}
A.~Khraisat, I.~Gondal, P.~Vamplew, and J.~Kamruzzaman.
\newblock Survey of intrusion detection systems: techniques, datasets and
  challenges.
\newblock {\em Cybersecurity}, 2(1):20, 2019.

\bibitem{survey_pure5}
A.~Aldweesh, A.~Derhab, and A.~Emam.
\newblock Deep learning approaches for anomaly-based intrusion detection
  systems: A survey, taxonomy, and open issues.
\newblock {\em Knowledge-based Systems}, 189:105--124, 2020.

\bibitem{survey_pure6}
G.~Fernandes, J.~J. P.~C. Rodrigues, L.~F. Carvalho, J.~F. Al-Muhtadi, and
  M.~L. Proença.
\newblock A comprehensive survey on network anomaly detection.
\newblock {\em Telecommunication Systems}, 70(3):447--489, 2019.

\bibitem{survey_pure7}
A.~L. {Buczak} and E.~{Guven}.
\newblock A survey of data mining and machine learning methods for cyber
  security intrusion detection.
\newblock {\em IEEE Communications Surveys Tutorials}, 18(2):1153--1176, 2016.

\bibitem{survey_pure8}
L.~N. {Tidjon}, M.~{Frappier}, and A.~{Mammar}.
\newblock Intrusion detection systems: A cross-domain overview.
\newblock {\em IEEE Communications Surveys Tutorials}, 21(4):3639--3681, 2019.

\bibitem{azwar}
H.~{Azwar}, M.~{Murtaz}, M.~{Siddique}, and S.~{Rehman}.
\newblock Intrusion detection in secure network for cybersecurity systems using
  machine learning and data mining.
\newblock In {\em 2018 IEEE 5th International Conference on Engineering
  Technologies and Applied Sciences (ICETAS)}, pages 1--9, 2018.

\bibitem{moustafa_table}
N.~{Moustafa}, J.~{Hu}, and J.~{Slay}.
\newblock A holistic review of network anomaly detection systems: A
  comprehensive survey.
\newblock {\em Journal of Network and Computer Applications}, 128:33--55, 2019.

\bibitem{Meena}
G.~{Meena} and R.~R. {Choudhary}.
\newblock A review paper on {IDS} classification using {KDD}99 and {NSL-KDD}
  dataset in {W}eka.
\newblock In {\em 2017 International Conference on Computer, Communications and
  Electronics (Comptelix)}, pages 553--558, 2017.

\bibitem{kdd_rama}
R.~Ravipati and A.~Munther.
\newblock A survey on different machine learning algorithms and weak
  classifiers based on {KDD} and {NSL-KDD} datasets.
\newblock {\em International Journal of Artificial Intelligence \&
  Applications}, 10:01--11, 2019.

\bibitem{kdd_rnn}
C.~{Yin}, Y.~{Zhu}, J.~{Fei}, and X.~{He}.
\newblock A deep learning approach for intrusion detection using recurrent
  neural networks.
\newblock {\em IEEE Access}, 5:21954--21961, 2017.

\bibitem{bengio}
H.~Bengio.
\newblock Learning deep architectures for {AI}.
\newblock {\em Found. Trends Mach. Learn.}, 2(1):1--127, 2009.

\bibitem{glorot}
X.~Glorot, A.~Bordes, and Y.~Bengio.
\newblock Deep sparse rectifier neural networks.
\newblock In {\em Proceedings of the Fourteenth International Conference on
  Artificial Intelligence and Statistics}, pages 315--323, 2011.

\bibitem{relu1}
G.~E. {Dahl}, D.~{Yu}, L.~{Deng}, and A.~{Acero}.
\newblock Context-dependent pre-trained deep neural networks for
  large-vocabulary speech recognition.
\newblock {\em IEEE Transactions on Audio, Speech, and Language Processing},
  20(1):30--42, 2012.

\bibitem{relu2}
M.~{Ravanelli}, P.~{Brakel}, M.~{Omologo}, and Y.~{Bengio}.
\newblock Light gated recurrent units for speech recognition.
\newblock {\em IEEE Transactions on Emerging Topics in Computational
  Intelligence}, 2(2):92--102, 2018.

\bibitem{softmax}
H.~{Le}, I.~{Oparin}, A.~{Allauzen}, J.~{Gauvain}, and F.~{Yvon}.
\newblock Structured output layer neural network language models for speech
  recognition.
\newblock {\em IEEE Transactions on Audio, Speech, and Language Processing},
  21(1):197--206, 2013.

\bibitem{cnn}
Y.~{Lecun} and Y.~{Bengio}.
\newblock Convolutional networks for images, speech and time series.
\newblock {\em The {MIT} Press}, pages 255--258, 1995.

\bibitem{rnn}
D.E. {Rumelhart}, G.E. {Hinton}, and R.J. {Williams}.
\newblock Learning representations by back-propagating errors.
\newblock {\em Nature}, 323(6088):533--536, 1986.

\bibitem{lstm}
S.~{Hochreiter} and J.~{Schmidhuber}.
\newblock Long {S}hort-{T}erm {M}emory.
\newblock {\em Neural Computation}, 9(8):1735--1780, 1997.

\bibitem{gru}
K.~Cho, B.~van Merrienboer, {\c{C}}.~G{\"{u}}l{\c{c}}ehre, D.~Bahdanau,
  F.~Bougares, H.~Schwenk, and Y.~Bengio.
\newblock Learning phrase representations using {RNN} encoder-decoder for
  statistical machine translation.
\newblock In {\em Proceedings of the 2014 Conference on Empirical Methods in
  Natural Language Processing}, pages 1724--1734, 2014.

\bibitem{wisard}
P.~Bowden I.~Alexander, W.~Thomas.
\newblock Learning deep architectures for ai.
\newblock {\em Sensor Review}, 4(3):120--124, 1984.

\bibitem{wisard_new}
M.~De Gregorio and M.~Giordano.
\newblock An experimental evaluation of weightless neural networks for
  multi-class classification.
\newblock {\em Applied Soft Computing}, 72:338 -- 354, 2018.

\bibitem{lvq}
T.~Kohonen.
\newblock Learning vector quantization.
\newblock In {\em The {H}andbook of {B}rain {T}heory and {N}eural {N}etworks
  ({A}rbib ed.)}, pages 631--635, 2003.

\bibitem{kohonen}
T.~Kohonen.
\newblock {\em Self-Organizing Maps, 3rd ed.}
\newblock Springer-Verlag, Berlin, Heidelberg, 2001.

\bibitem{aggarwal}
C.~C. Aggarwal.
\newblock {\em Neural Networks and Deep Learning}.
\newblock Springer-Verlag, Gewerbestrasse 11, 6330 Cham, Switzerland, 2018.

\bibitem{adam}
D.P. Kingma and L.J. Ba.
\newblock Adam: A method for stochastic optimization.
\newblock In {\em ICLR2015}, 2015.

\bibitem{ddos1bis}
V.~{Matta}, M.~{Di Mauro}, and M.~{Longo}.
\newblock Botnet identification in randomized {DD}o{S} attacks.
\newblock In {\em Proceedings of the 24th European Signal Processing
  Conference}, pages 2260--2264, 2016.

\bibitem{ddos2}
M.~E. {Ahmed}, S.~{Ullah}, and H.~{Kim}.
\newblock Statistical application fingerprinting for {DD}o{S} attack
  mitigation.
\newblock {\em IEEE Transactions on Information Forensics and Security},
  14(6):1471--1484, 2019.

\bibitem{accuracy2}
J.~{Zhang}, Y.~{Xiang}, Y.~{Wang}, W.~{Zhou}, Y.~{Xiang}, and Y.~{Guan}.
\newblock Network traffic classification using correlation information.
\newblock {\em IEEE Transactions on Parallel and Distributed Systems},
  24(1):104--117, 2013.

\bibitem{tnsm-perf}
A.~{Alsirhani}, S.~{Sampalli}, and P.~{Bodorik}.
\newblock Ddos detection system: Using a set of classification algorithms
  controlled by fuzzy logic system in {A}pache {S}park.
\newblock {\em IEEE Transactions on Network and Service Management},
  16(3):936--949, 2019.

\bibitem{lisboa}
P.~S.~Szczepaniak P.~J.G.~Lisboa, E. C.~Ifeachor.
\newblock {\em Artificial Neural Networks in Biomedicine}.
\newblock Springer Science, London, 2000.

\bibitem{prasad08}
V.~Prasad and S.D. Gupta.
\newblock {\em Applications And Potentials Of Artificial Neural Networks In
  Plant Tissue Culture}.
\newblock Springer, Dordrecht, 2008.

\bibitem{cnn_fluct}
S.J. Kim, C.~Wang, B.~Zhao, H.~Im, J.~Min, H.J. Choi, J.~Hee, J.~Tadros, N.T.
  Choi, C.M. Castro, R.~Weissleder, H.~Lee, and K.~Lee.
\newblock Deep transfer learning-based hologram classification for molecular
  diagnostics.
\newblock {\em Scientific Reports}, 8:01--12, 2018.

\end{thebibliography}

\atColsBreak{\vskip 10pt}

\begin{IEEEbiography}[{\includegraphics[width=1in,height=1.13in,clip,keepaspectratio]{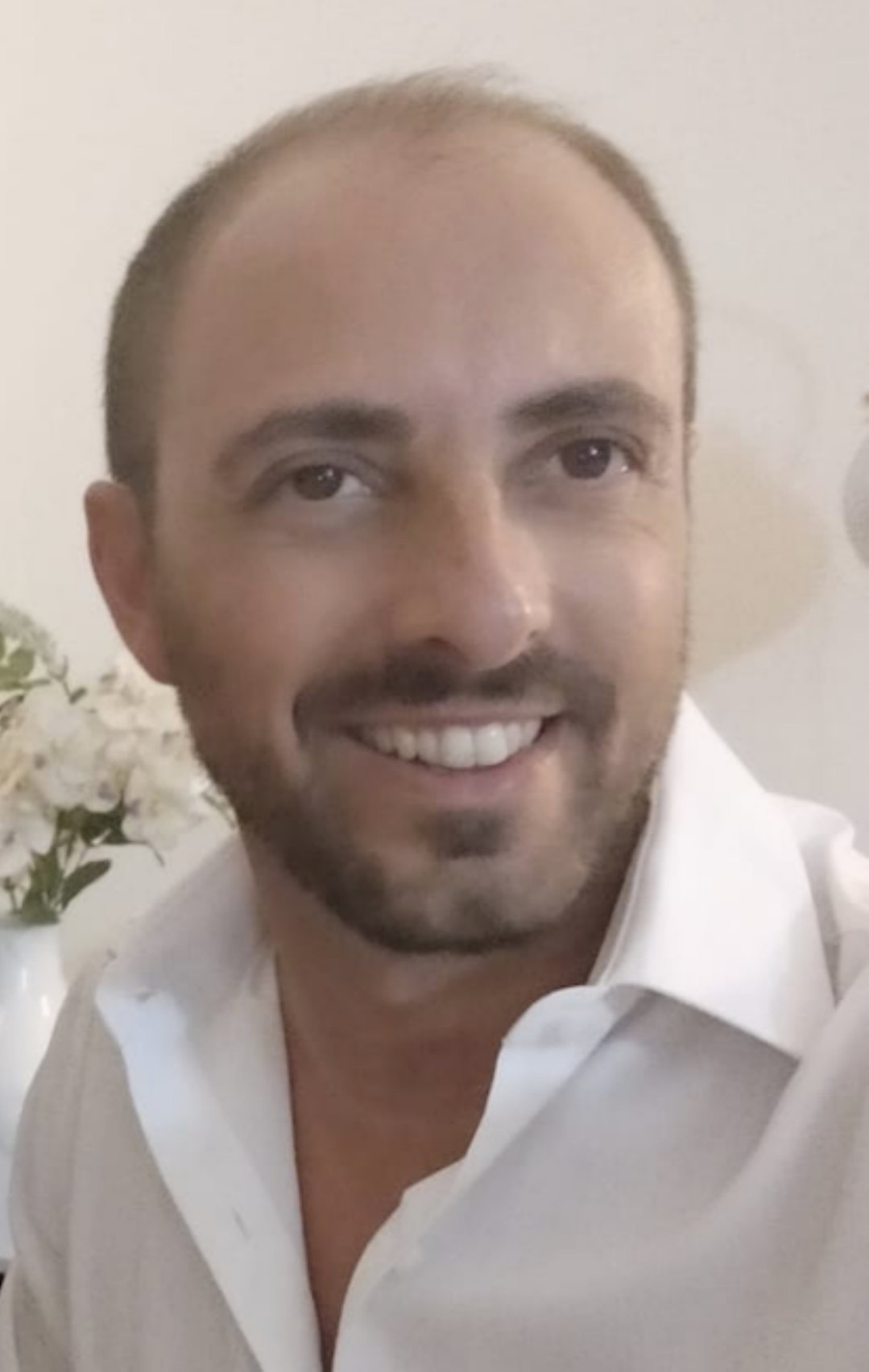}}] {Mario Di Mauro} received the Laurea degree in electronic engineering from the University of Salerno (Italy) in 2005, the M.S. degree in networking from the University of L'Aquila (Italy) jointly with the Telecom Italia Centre in 2006, and the PhD. degree in information engineering in 2018 from University of Salerno.
He was a Research Engineer with CoRiTel (Research Consortium on Telecommunications, led by Ericsson Lab, Italy) and then a Research Fellow with University of Salerno. He has authored several scientific papers, and holds a patent on a telecommunication aid for impaired people. His main fields of interest include: network security, performance and availability, data analysis, ML techniques. 
\end{IEEEbiography}

%\vspace{-50mm}
\begin{IEEEbiography}[{\includegraphics[width=1in,height=1.13in,clip,keepaspectratio,angle=270]{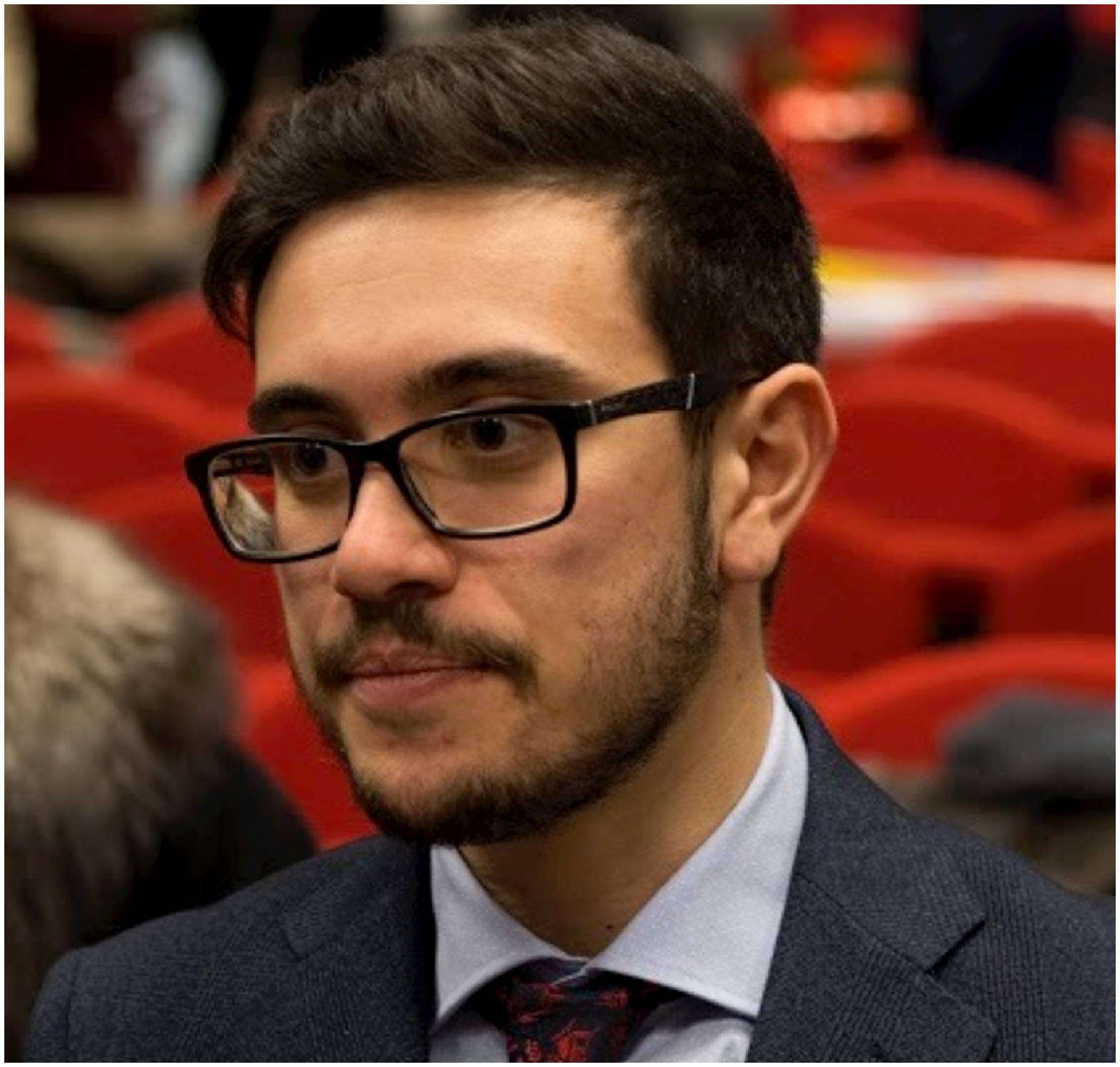}}] {Giovanni Galatro} received the Laurea degree (summa cum laude) in information engineering from the University of Salerno (Italy) in 2018, and has been a visiting student at Dept. of Computer Science at Groningen University (Netherlands). In 2017 he got a scholarship with Telecommunication and Applied Statistics groups, focused on the availability analysis of modern telecommunication systems. His main fileds of interest include: network availability and machine learning.
\end{IEEEbiography}

%\vspace{-50mm}
\begin{IEEEbiography}[{\includegraphics[width=1in,height=1.13in,clip,keepaspectratio]{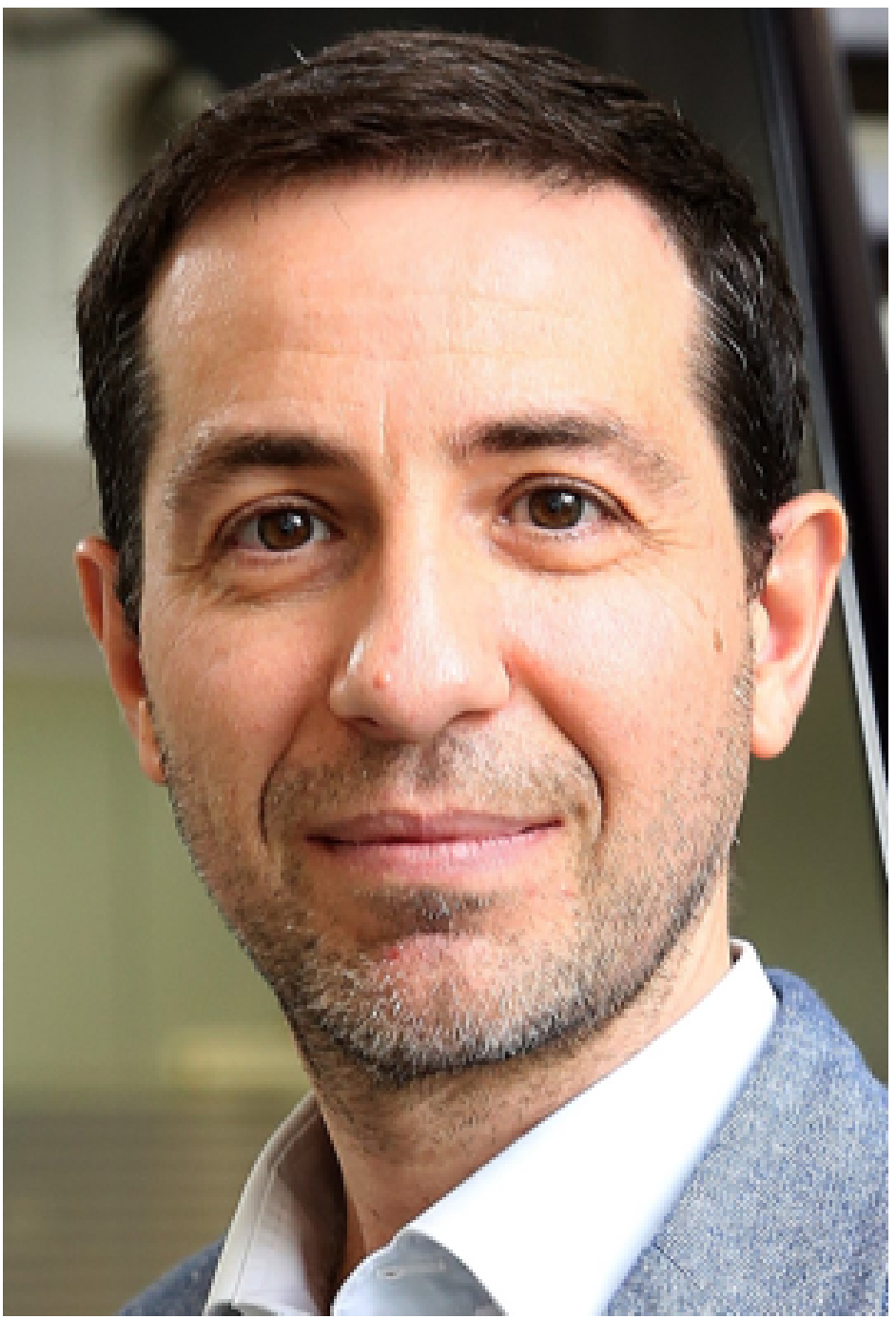}}]  {Antonio Liotta} (SMEEE’15) is Full Professor at the Faculty of Computer Science, Free University of Bozen-Bolzano (Italy), where he teaches Data Science and Computer Networks. Previously, he was full professor at Eindhoven University of Technology (NL), University of Derby (UK), and Edinburgh Napier University (UK).  He has also held academic positions at University of Surrey (UK) and Essex University (UK), in addition to Visiting and Distinguished Professorships in the UK, Australia and China. His team is at the forefront of influential research in data science and artificial intelligence, specifically in the context of smart cities, Internet of Things, and smart sensing. He is renowned for his contributions to miniaturized machine learning, particularly in the context of the Internet of Things. He has led the international team that has recently made a breakthrough in artificial neural networks, using network science to accelerate the training process. Antonio is a Fellow of the U.K. Higher Education Academy. He has 6 patents and over 350 publications to his credit, and is the Editor-in-Chief of the Springer Internet of Things book series.
\end{IEEEbiography}

\end{document}